\documentclass[%
 amsmath,amssymb,
]{revtex4-1}

\usepackage{graphicx}
\usepackage{dcolumn}
\usepackage{bm}
\usepackage{hyperref}

\usepackage{float}

\usepackage{enumerate}

\begin{document}


\title{Long-term fluctuations in globally coupled phase oscillators with general coupling: Finite size effects}

\author{Isao Nishikawa$^{1,2}$, Gouhei Tanaka$^{1,2}$,
Takehiko Horita$^{3}$, and Kazuyuki Aihara$^{1,2}$}
\affiliation{
$^1$Department of Mathematical Informatics, Graduate School of Information Science and Technology, University of Tokyo, 7-3-1 Hongo, Bunkyo-ku, Tokyo 113-8656, Japan
}%
\affiliation{
$^{2}$Institute of Industrial Science, University of Tokyo, Tokyo 153-8505, Japan
}%
\affiliation{
$^{3}$Department of Mathematical Sciences, Osaka Prefecture University,
Sakai 599-8531, Japan
}%




\date{\today}

\begin{abstract}
We investigate the diffusion coefficient of the time integral of the Kuramoto order parameter in globally coupled nonidentical phase oscillators.
This coefficient represents the deviation of the time integral of the order parameter from its mean value on the sample average.
In other words,
this coefficient characterizes long-term fluctuations of the order parameter.
For a system of $N$ coupled oscillators,
we introduce a statistical quantity $D$, which denotes the product of $N$ and the diffusion coefficient.
We study the scaling law of $D$ with respect to the system size $N$.
In other well-known models such as the Ising model,
the scaling property of $D$ is $D \sim O(1)$ for both coherent and incoherent regimes except for the transition point.
In contrast, in the globally coupled phase oscillators, the scaling law of $D$ is different for the coherent and  incoherent regimes:
$D \sim O(1/N^a)$ with a certain constant $a>0$ in the coherent regime, and $D \sim O(1)$ in the incoherent regime.
We demonstrate that these scaling laws hold for
several representative coupling schemes.
\begin{description}
\item[PACS numbers]
05.45.Xt, 05.50.+q, 05.40.-a, 05.60.-k

\end{description}
\end{abstract}

\pacs{Valid PACS appear here}
\maketitle



\textbf{In real-world systems,
large populations of coupled oscillators often experience global synchronous oscillations.
To elucidate the general properties of such phenomena,
considerable research has been conducted on simple models of globally coupled
phase oscillators.
The Kuramoto order parameter has been widely used to measure the degree of synchronization
and to characterize the synchronization transition in the phase oscillator model.
When the system size is infinite,
fluctuations of the order parameter vanish after a transient period;
the scaling law for this parameter
has been well studied for a general coupling function.
However, when the system is large but of finite size,
the fluctuations in the order parameter do not vanish and
their behavior has not been fully understood.
It is not clear
whether the conventional standard statistical quantities such as the variance and the correlation time of the order parameter
can fully characterize its fluctuation behavior.
Further, the dependence of the statistical properties of the order parameter on the coupling scheme
is still not completely understood.
As a step toward understanding these problems,
we focus on a statistical quantity that characterizes long-term fluctuations in the order parameter.
In other well-known models such as the Ising model,
the scaling property of the statistical quantity with respect to the system size is the same for coherent and incoherent regimes except for the transition point.
In contrast, in the globally coupled phase oscillators,
the decay speed of the statistical quantity in the coherent regime is faster than that in the incoherent regime.
This difference is caused by a difference in the correlations among the phases of the oscillators at different times.
We show that the scaling laws hold for a large class of general coupling schemes.
}

\section{Introduction}

Nonlinear systems are often used for modeling chemical reactions, engineering circuits, and biological populations \cite{PikovskyBook}.
Synchronization in such systems has attracted considerable attention in the past several decades.
The phase description of the systems is one of the most effective methods to understand synchronization in interacting oscillatory systems \cite{PikovskyBook,KuramotoBook}.
Accordingly, there have been a number of studies on the globally coupled phase oscillator model \cite{KuramotoBook}, which is described as follows:
\begin{align}
\label{phasemodel}
\dot \theta_j = \omega_j + \frac{K}{N} \sum _{k=1} ^N h(\theta_k - \theta _j), \ (j=1,\ldots,N),
\end{align}
where $\theta _j$ represents the phase of the $j$th oscillator,
$\omega_j$ represents the natural frequency of the $j$th oscillator,
$K>0$ represents the coupling strength,
$h$ is the coupling function,
and $N$ is the number of oscillators.
The oscillators are synchronized when the coupling strength is sufficiently large for the Kuramoto model where $h(x) = \sin(x)$ \cite{KuramotoBook,Kuramotomodel}.
The synchronization transition in the phase oscillator model $(\ref{phasemodel})$ with infinite dimension (i.e., in the thermodynamic limit $N\rightarrow \infty $)
has been well studied with regard to its analogy to the second-order phase transition \cite{KuramotoBook,Kuramotomodel,Sakaguchi2,Daido2,Crawford2,Chiba2,Chiba}.
In particular, one of the main areas of focus in these studies
has been the behavior of the order parameter $R(t)$ as a measure of synchrony \cite{KuramotoBook},
which is defined as follows:
\begin{align}
\label{R}
R(t) \equiv \frac {1}{N} \left|\sum _{j=1} ^N \exp(2\pi i\theta _j)\right|.
\end{align}
A synchronization transition can be characterized by a change in the order parameter from zero to a non-zero value with an increase of $K$.
We assume that the stationary state with $R(t)=0$ in the incoherent (desynchronized) regime supercritically bifurcates at the critical coupling strength $K=K_c$,
above which the oscillators are synchronized or coherent.
The scaling property of the order parameter around the synchronization transition point has been intensively studied:
first, Kuramoto \cite{KuramotoBook} analytically investigated
the phase oscillator model $(\ref{phasemodel})$ with the sinusoidal coupling function; 
then, Daido \cite{Daido2}, Crawford and Davies \cite{Crawford2}, and Chiba and Nishikawa \cite{Chiba2}
considered the phase oscillator model $(\ref{phasemodel})$ with more general coupling functions.

However, finite size effects in the phase oscillator model $(\ref{phasemodel})$ have not yet been fully understood.
It is not clear
whether the conventional standard statistical quantities such as the variance and the correlation time of the order parameter
can fully reveal the characteristics of its fluctuations.
Further, the dependence of
the statistical properties of the order parameter on the coupling scheme is still not completely understood,
because most previous studies on the finite size effects have only examined
the Kuramoto model \cite{Daido,Daido4,Pikovsky4,Hong,Hong3}.
It is known that the scaling law of the order parameter in the infinite-size system depends on
the coupling function \cite{Daido2,Crawford2}.
However, it is not evident how the coupling function influences the scaling law of the order parameter in a finite-size system.
Although the finite size effects in the incoherent regime were addressed in the case of general coupling \cite{Hildebrand,Buice},
those in the coherent regime are still unclear.

This paper investigates the finite size effects on the long-term fluctuations of the order parameter
in the phase oscillator model $(\ref{phasemodel})$ by using the diffusion coefficient of the time integral of the order parameter.
This diffusion coefficient represents the deviation of the time integral of the order parameter from its mean value on the sample average.
Although this statistical quantity has been used in the large deviation theory,
it has not been examined in the literature of coupled phase oscillators.
Denoting the product of $N$ and the diffusion coefficient as $D$,
we analyze the properties of $D$ in the phase oscillator models with the sinusoidal coupling function
and also with more general coupling functions.

\begin{figure}
\centering
\begin{tabular}{ll}
\includegraphics[width=3.5cm,clip]{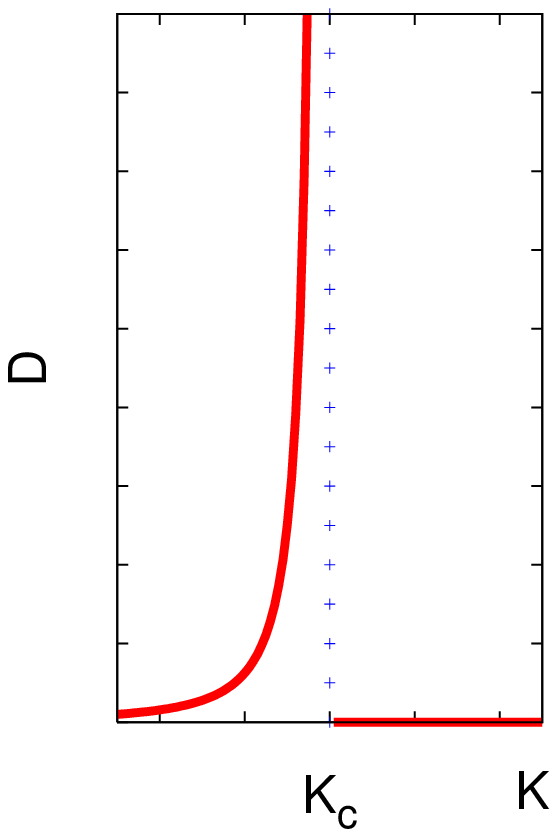}
\includegraphics[width=3.5cm,clip]{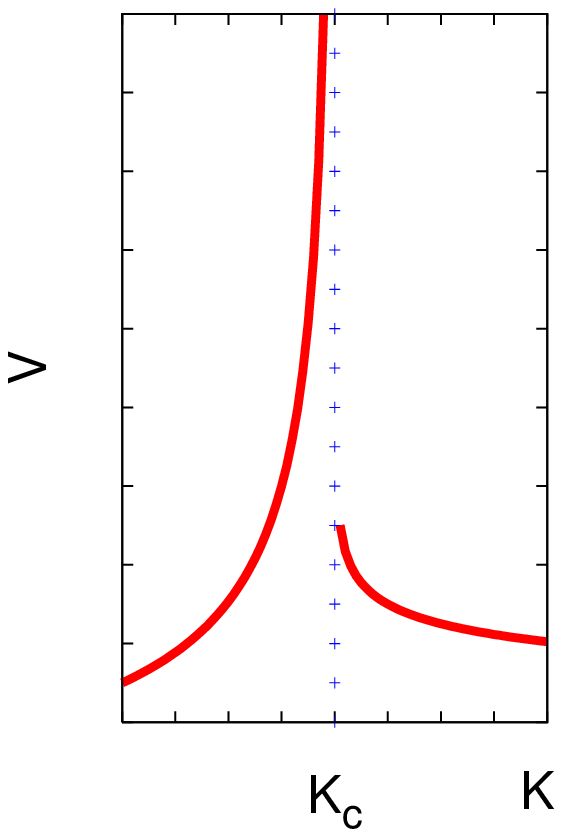}
\end{tabular}
\caption{\label{fig:schematic}
Schematic view of $D$ and $V$ in the limit $N \rightarrow \infty $
when the synchronization transition at $K=K_c$ is similar to the second order phase transition.
}
\end{figure}

We show that the scaling law of $D$ with respect to system size $N$ is different
for the coherent and incoherent regimes in the model $(\ref{phasemodel})$ with a general coupling function:
$D \sim O(1/N^a)$ with a certain positive constant $a$ in the coherent regime,
and $D \sim O(1)$ in the incoherent regime.
The scaling law in the coherent regime is anomalous
because the scaling law of $D$ is $D \sim O(1)$ for both the two regimes in other well-known systems such as the Ising model.
Moreover, we analytically demonstrate that in the coherent regime, $D=0$ in the limit $N \rightarrow \infty $.
Therefore, the statistical quantity $D$ is useful to
qualitatively differentiate between the coherent and incoherent regimes, as illustrated in Fig.~$\ref{fig:schematic}$.
This property is not found in other statistical quantities such as the variance of the
order parameter.
When we denote the product of $N$ and this variance by $V$,
it follows $V \sim O(1)$ with system size $N$ and $V\not =0$ in the limit $N \rightarrow \infty $ in both
coherent and incoherent regimes \cite{Daido}, as shown in Fig.~$\ref{fig:schematic}$.

\section{A statistical quantity $D$ characterizing long-term fluctuations of the order parameter}

We introduce the diffusion coefficient of the time integral of $R(t)$ to characterize the long-term fluctuations of $R(t)$.
The variance of the time integral $\int _0 ^t R(s) ds$ is given as follows:
\begin{align}
\label{sigma}
\sigma^2(t) \equiv N \left[ \left\langle \left( \int _{t_0} ^{t+t_0} R(s)ds - \langle R \rangle _t  t \right) ^2 \right\rangle _t \right] _s,
\end{align}
where $\langle \cdot \rangle _t$ and $[ \cdot ] _s$ represent 
the time average over the period from $t_0=0$ to $t_0 \rightarrow \infty $ and
the sample average of different realizations of $\omega _j $ which are independently chosen from a certain distribution, respectively.
Then, the following diffusion law holds:
\begin{align}
\label{D}
D \equiv \lim _{t\rightarrow \infty } \sigma^2(t) /  2t,
\end{align}
which represents the deviation of $\int _0 ^t R(s)ds$ from its mean value $\langle R \rangle _t  t$ on the sample average.

It should be noted that the statistical quantity $D$ is different from the variance of $R(t)$, given by
\begin{align}
\label{V}
V \equiv   N  [ \langle ( R(t_0) - \langle R \rangle _t)^2 \rangle _t ] _s,
\end{align}
which characterizes instantaneous fluctuations of $R(t)$.
For example, if $R(t)$ oscillates periodically, then $D = 0$ whereas $V>0$.

Now, we consider the dependence of $D$ on other statistical quantities.
We define the correlation function of $R(t)$ as follows:
\begin{align}
\label{C}
C(t) \equiv N  [ \langle ( R(t+t_0) - \langle R \rangle _t)( R(t_0) - \langle R \rangle _t) \rangle _t ] _s.
\end{align}
Because $C(0) = V$, we obtain
\begin{align}
\label{CC}
C(t) = V  f(t/\tau ),
\end{align}
where $\tau $ and $f$ with $f(0) = 1$ represent the correlation time and the normalized correlation function, respectively.
In this paper, $\tau $ is defined as the value satisfying $C(\tau) = C(0)/2$.
Then, $D$ satisfies the following equation \cite{Correlationfunction}:
\begin{align}
\label{CD}
D = \frac{1}{2} \int _ {-\infty} ^{\infty} C(s) ds.
\end{align}
Therefore, from Eqs.~($\ref{CC}$)-($\ref{CD}$), $D$ can be described using $V$, $\tau$, and $f$ as follows:
\begin{align}
\label{DD}
D = \frac{1}{2}  V  \tau  \int _{-\infty } ^{\infty } f(s)ds.
\end{align}

\section{Scaling law of $D$ with system size for the Kuramoto model}

\begin{figure}
\centering
\begin{center}
\begin{tabular}{ll}
\includegraphics[width=6cm,clip]{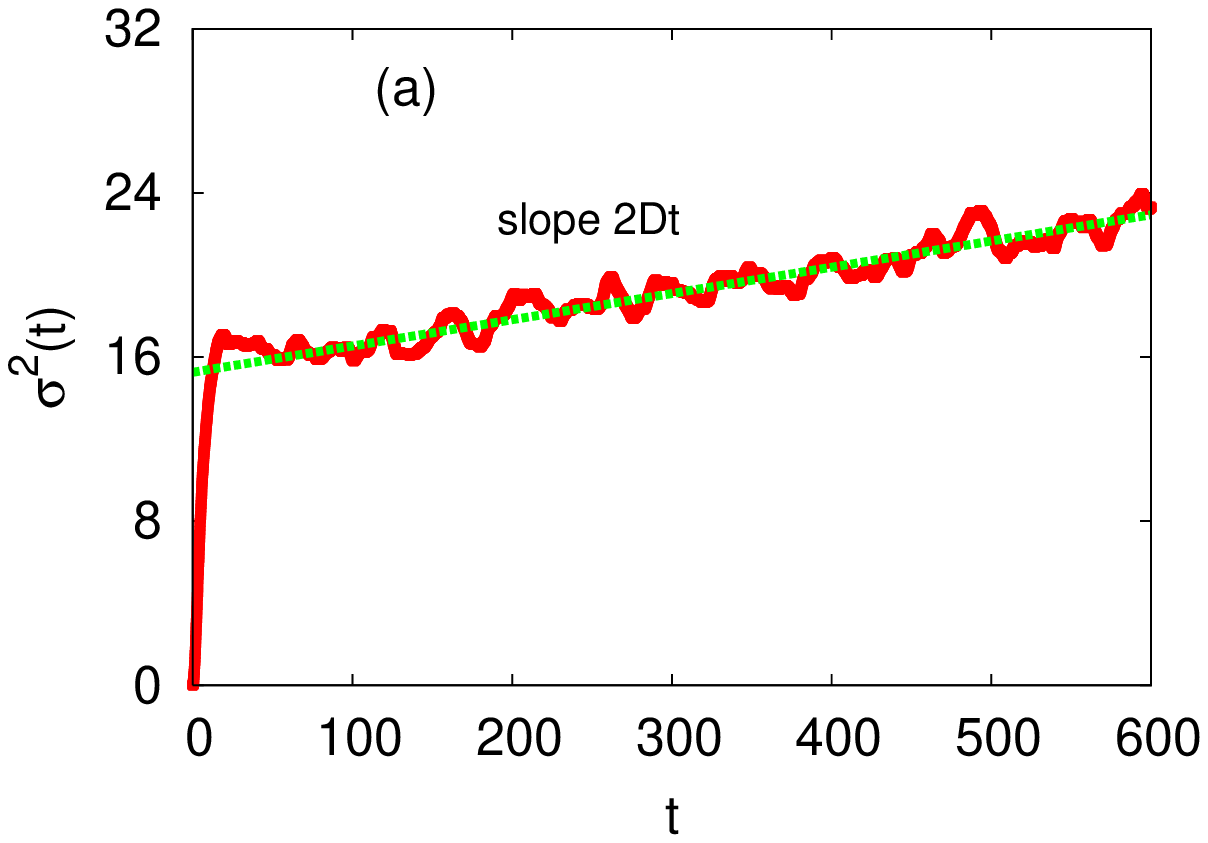}
\includegraphics[width=6cm,clip]{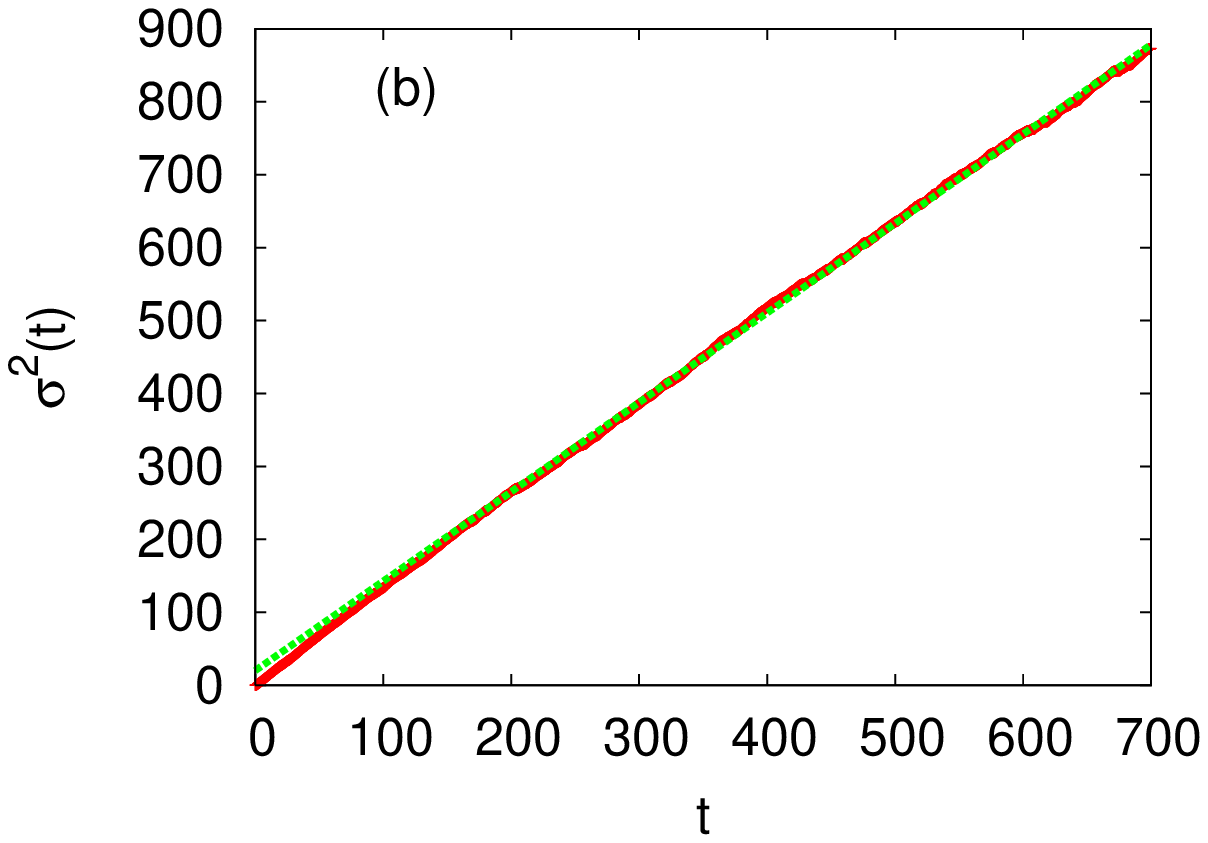}
\end{tabular}
\end{center}
\begin{center}
\begin{tabular}{ll}
\includegraphics[width=6cm,clip]{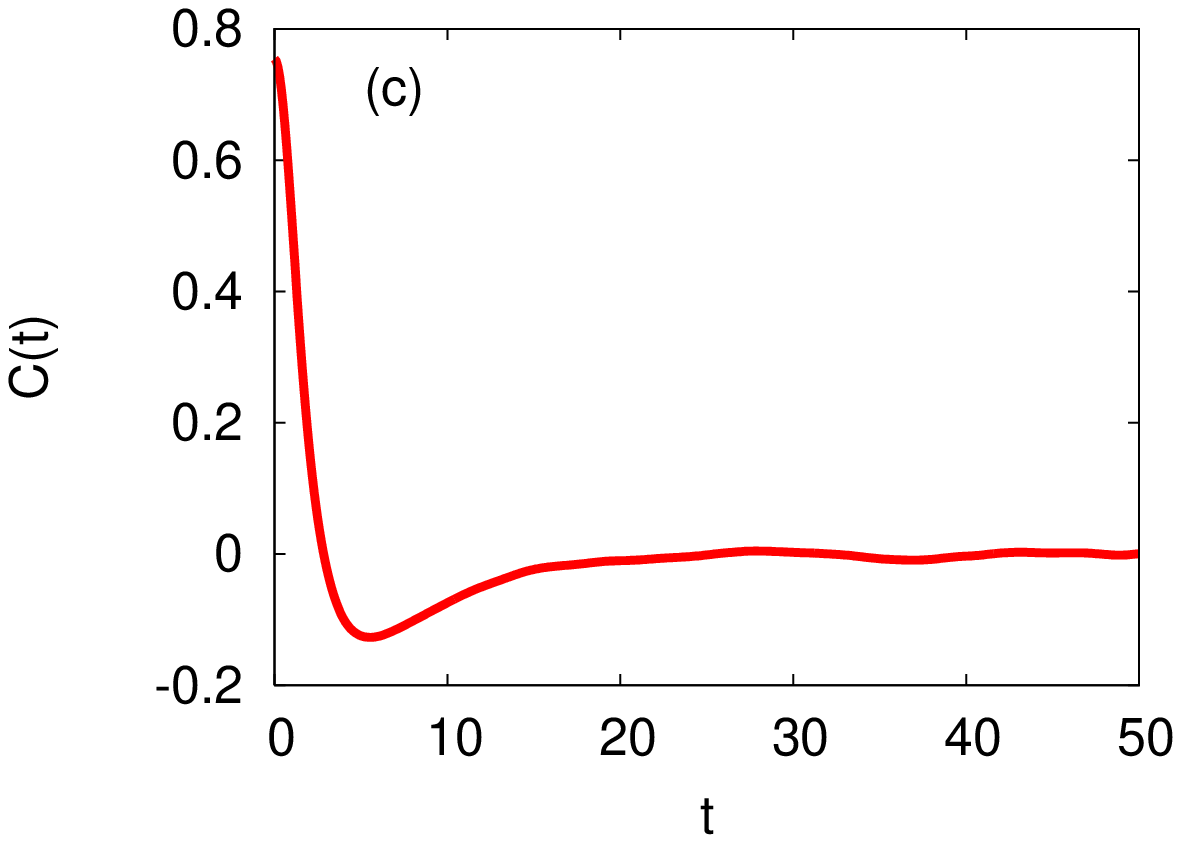}
\includegraphics[width=6cm,clip]{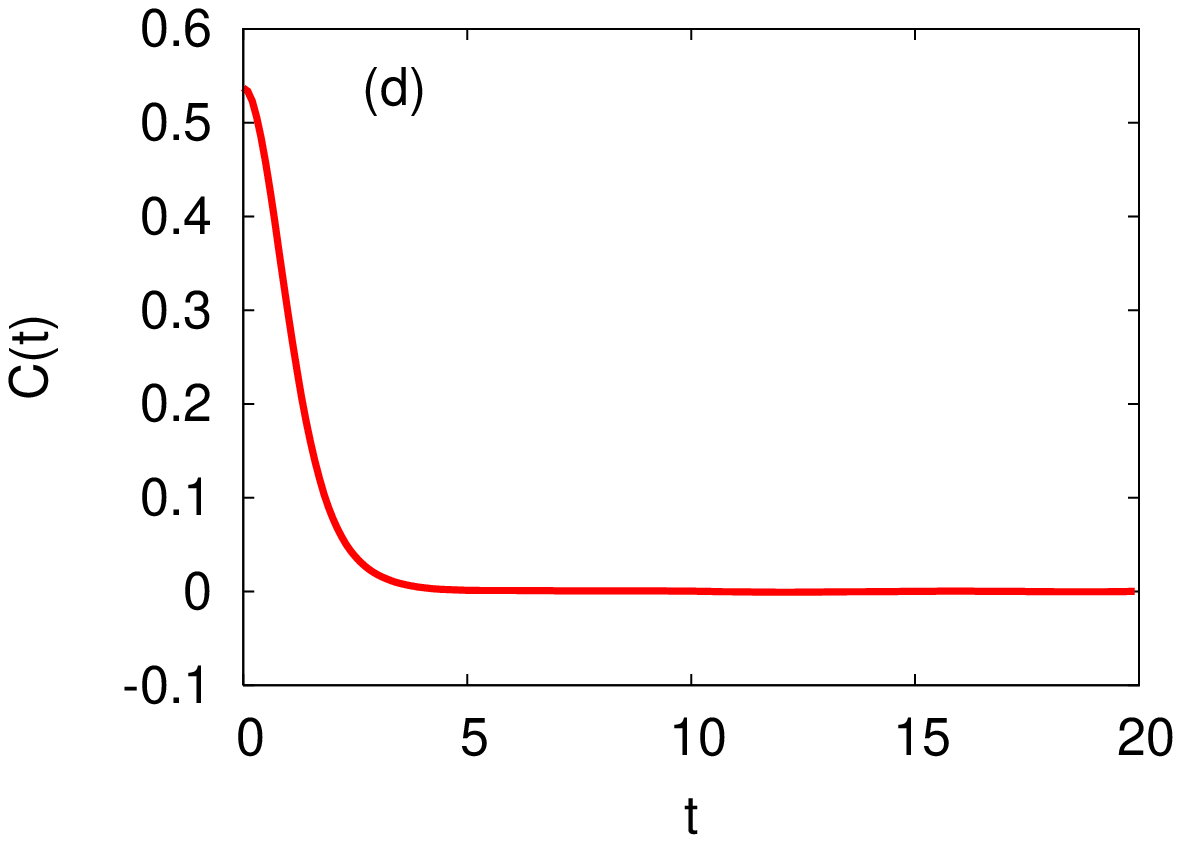}
\end{tabular}
\end{center}
\caption{The time evolutions of the variance $\sigma ^2 (t)$ of the integrated order parameter (upper) and the correlation function $C(t)$ (lower)
in the Kuramoto model where $h(x) = \sin(x)$ and $N=24000$.
(a)(c) The coherent regime where $K=1.68 > K_c = 1.59\cdots$.
(b)(d) The incoherent regime where $K = 0.8 < K_c$.
In both regimes, $\sigma ^2(t)$ increases linearly with slope $2Dt$ after a transient period.
The correlation function $C(t)$ in the coherent regime has a characteristic time period in which $C(t)<0$,
whereas the form of $C(t)$ in the incoherent regime is almost exponential.
Each plot is an average over 30 samples.
}
\label{fig:sigma}
\end{figure}

The main result obtained for the scaling law of $D$ is as follows:
in the Kuramoto model (Eq.~($\ref{phasemodel}$) with $h(x) = \sin (x)$),
the asymptotic form of $D$ for large $N$ takes
\begin{align}
\label{DN} D \sim
\begin{cases}
O(1/N^a) & \mathrm{(coherent \ regime)},\\
O(1) & \mathrm{(incoherent \ regime)},
\end{cases}
\end{align}
where $a>0$ is a certain constant.

The cause of the difference in the scaling laws can be intuitively understood from Eq.~$(\ref{DD})$ as follows.
It is known that $V \sim O(1)$ and $\tau \sim O(1)$ for the phase oscillator model $(\ref{phasemodel})$  \cite{Daido,Hildebrand,Buice}
and also for other well-known models \cite{Goldenfeldbook,Nishimori}
except for the synchronization transition point.
If $f(s)$ is written in the simple exponential form, as is commonly the case \cite{Goldenfeldbook,Nishimori},
then $\int _{-\infty } ^{\infty } f(s)ds$ is finite.
In fact, the form of $f(s)$ is almost exponential in the incoherent regime of the Kuramoto model as numerically shown later.
Therefore, with consideration of the above facts,
we can infer $D \sim O(1)$ from Eq.~$(\ref{DD})$ in the incoherent regime.
However, in the coherent regime, $f(s)$ is not in a simple exponential form \cite{Daido},
and thereby the scaling law of $D$ is different from $D \sim O(1)$.

This section shows the scaling law $(\ref{DN})$ numerically.
We assume that the distribution of the natural frequencies $\omega _j$ is Gaussian with mean zero and variance one.
In numerical simulations,
the natural frequencies are generated from the Gaussian distribution in a random manner.
Figure~$\ref{fig:sigma}$ shows the differences in the time evolutions of $\sigma ^2(t)$ in Eq.~$(\ref{sigma})$ and $C(t)$ in Eq.~$(\ref{C})$ between the coherent and incoherent regimes.
The value of $D$ is estimated by fitting the values of $\sigma ^2(t)$ with a line of slope $2Dt$ for a sufficiently large $t$.
We separately consider the coherent and incoherent regimes.

\subsection{Coherent regime}

\begin{figure}
\centering
\includegraphics[width=6cm,clip]{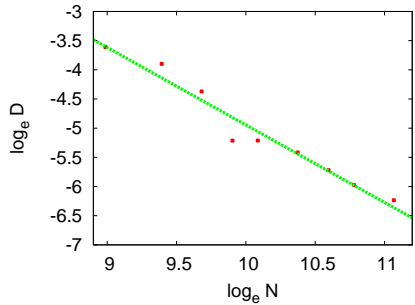}
\caption{The scaling property of the diffusion coefficient $D$ with system size $N$ in the coherent regime of the Kuramoto model,
where $K = 1.68 > K_c = 1.59\cdots$, and
$N= 8000,\ldots, 64000$.
Line fitting yields $D\sim N^{-1.33}$.
Each plot is an average over 90 samples.
}
\label{fig:VD}
\end{figure}

Figure $\ref{fig:VD}$ shows the dependence of $D$ on the system size $N$.
We clearly see that $D$ is scaled as $D\sim O(1/N^a)$ where $a=1.33$ for $K=1.68$.
In Sec. V, we analytically show that $D=0$ in the limit $N \rightarrow \infty $.

\subsection{Incoherent regime}

\begin{figure}
\centering
\includegraphics[width=6cm,clip]{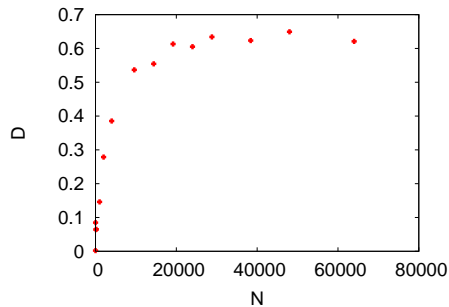}
\caption{The diffusion coefficient $D$ with an increase of system size $N$ in the incoherent regime of the Kuramoto model, where $K = 0.8 < K_c = 1.59\cdots$.
$D$ fluctuates around a finite value for sufficiently large $N$.
Each plot is an average over 90 samples.
}
\label{fig:D2}
\end{figure}

\begin{figure}
\centering
\begin{center}
\begin{tabular}{lll}
\includegraphics[width=5cm,clip]{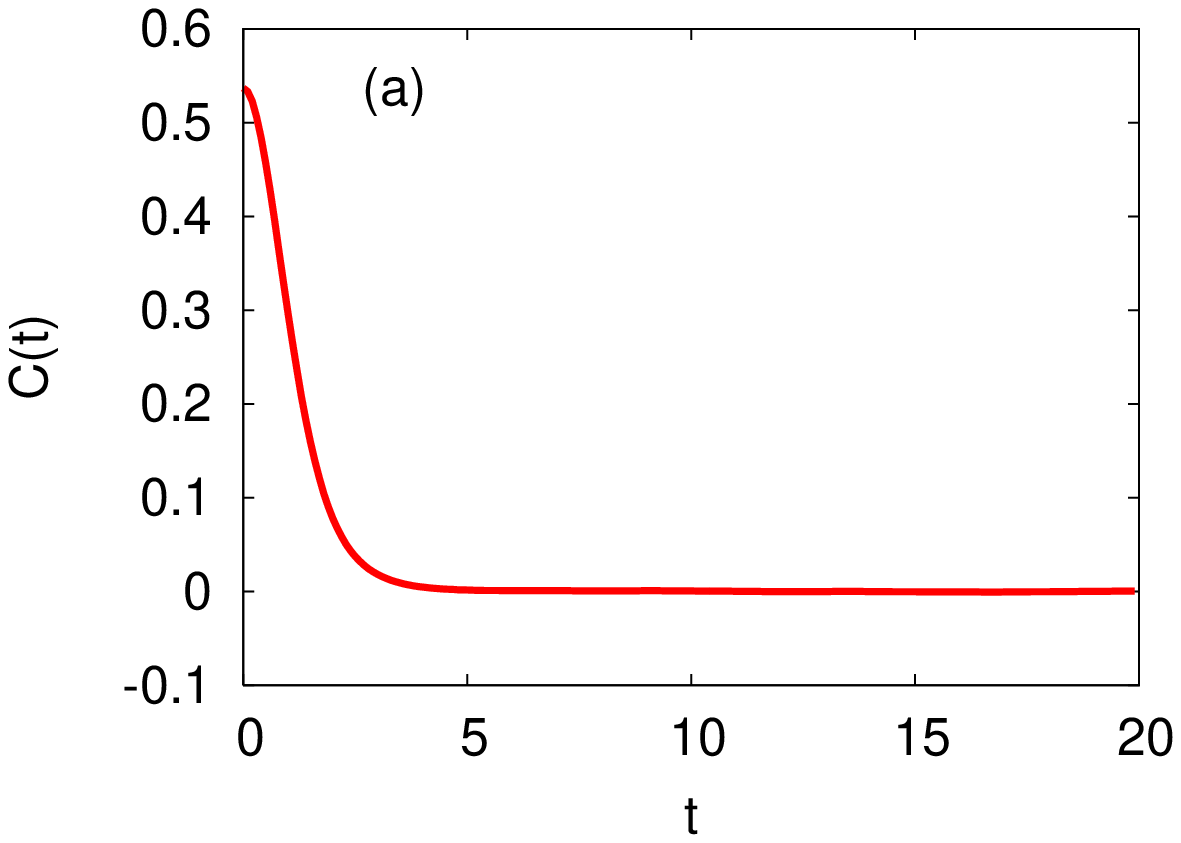}
\includegraphics[width=5cm,clip]{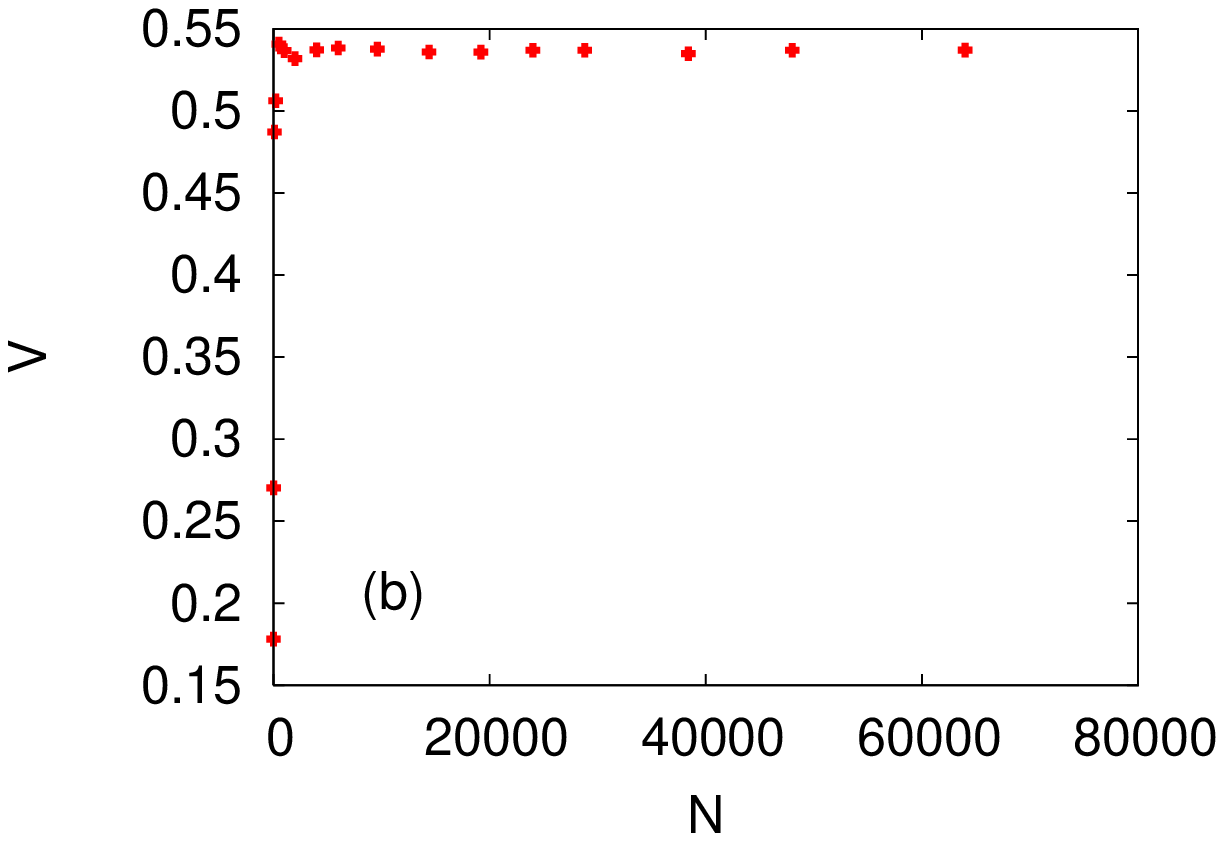}
\includegraphics[width=5cm,clip]{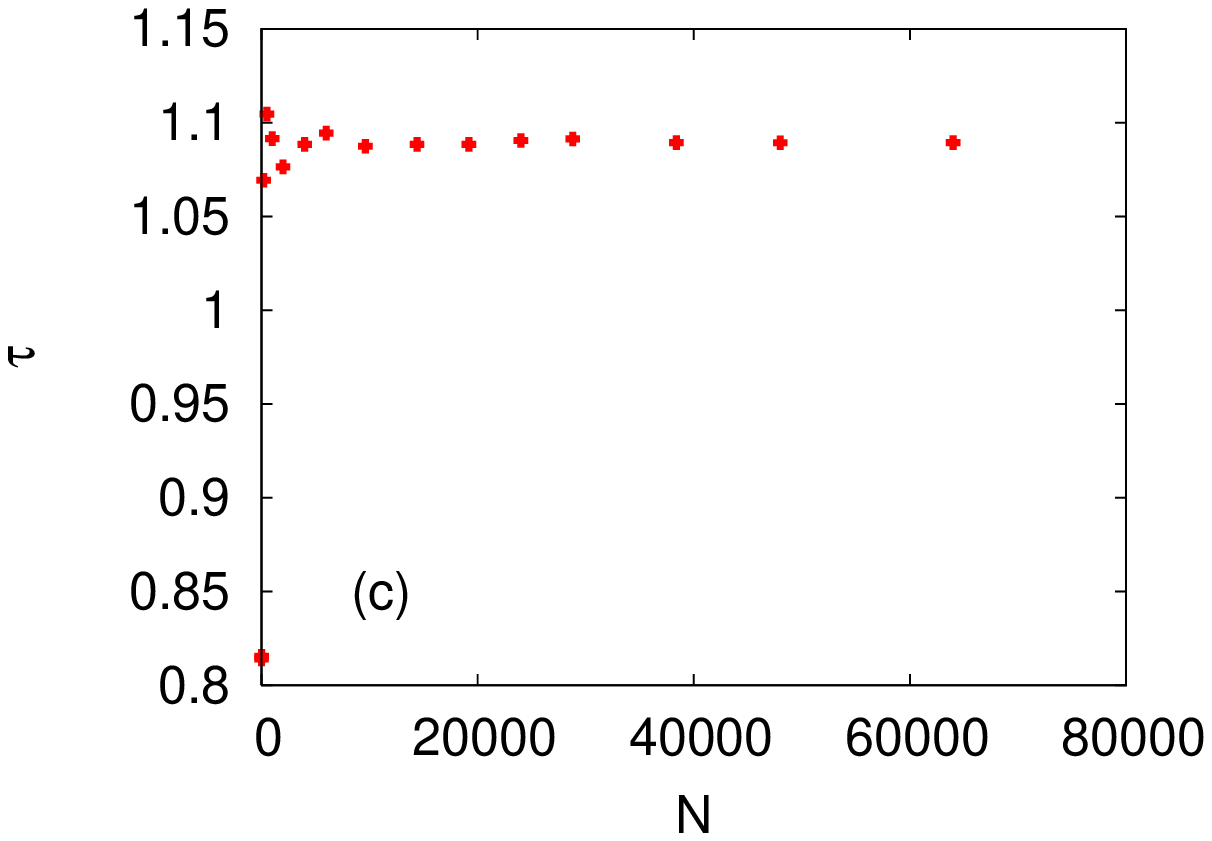}
\end{tabular}
\end{center}
\caption{The correlation function $C(t)$ for $N=64000$ (a), the variance $V$ (b), and the correlation time $\tau $ (c) of the order parameter
in the incoherent regime of the Kuramoto model, where $K = 0.8 < K_c = 1.59\cdots$.
$C(t)$ almost exponentially decreases with $t$.
$V$ and $\tau $ fluctuate around a finite value for sufficiently large $N$.
Each plot is an average over 90 samples.}
\label{fig:C}
\end{figure}

The diffusion coefficient $D$ fluctuates around a finite value for sufficiently large $N$
as shown in Fig.~$\ref{fig:D2}$.
It implies $D\sim O(1)$.
We support this scaling property by using Eq.~($\ref{DD}$).
The form of the correlation function $C(t)$ of the order parameter is almost exponential
as shown in Fig.~$\ref{fig:C}$(a) \cite{Daido}.
Therefore, if $V\sim O(1)$ and $\tau \sim O(1)$, $D$ should be of $O(1)$ from Eq.~($\ref{DD}$).
In fact, our numerical simulations confirm that
$V \sim O(1)$ \cite{Daido,Hildebrand,Buice} and $\tau \sim O(1)$ \cite{Daido,Hildebrand,Buice,Goldenfeldbook,Nishimori}
as shown in Figs.~$\ref{fig:C}$(b) and $\ref{fig:C}$(c).
Therefore, we conclude $D\sim O(1)$.

\section{Scaling law of $D$ with system size for more general couplings}

This section numerically confirms the scaling law $(\ref{DN})$ for other representative couplings to enhance the generality of our result.
The natural frequencies $\omega _j$ are chosen from the Gaussian distribution in a random manner as in the previous section.
Again we separately consider the coherent and incoherent regimes.

\subsection{Coherent regime}

In addition to the sinusoidal coupling function treated in the previous section,
we consider the following three coupling functions \cite{Crawford2}:
\begin{enumerate}[(i)]
\item A coupling of a generic form with nonzero second harmonic:

$h(x) = \sin(x) - (1/2) \sin(2x)$,
\item A coupling without the second harmonic term but with the third harmonic term:

$h(x) = \sin(x) - (1/2) \sin(3x)$,
\item A coupling without the symmetry:

$h(x) = \sin(x + \pi/4)$.
\end{enumerate}
We choose these coupling functions because
most coupling functions are classified on the basis of
the value of the critical exponent of the order parameter $R$
(for example, see p.30 and p.31 in \cite{Crawford2}) into the following three cases:
the coupling function is sinusoidal (Sec. III);
the coupling function has the second harmonic term \cite{Daido2,Crawford2} (case (i));
and
the coupling function lacks the second harmonic term and possesses the third harmonic term \cite{Crawford2} (case (ii)).
Additionally, case (iii) is considered to examine the effect of asymmetry in the coupling function.

\begin{figure}
\centering
\begin{center}
\begin{tabular}{lll}
\includegraphics[width=5cm,clip]{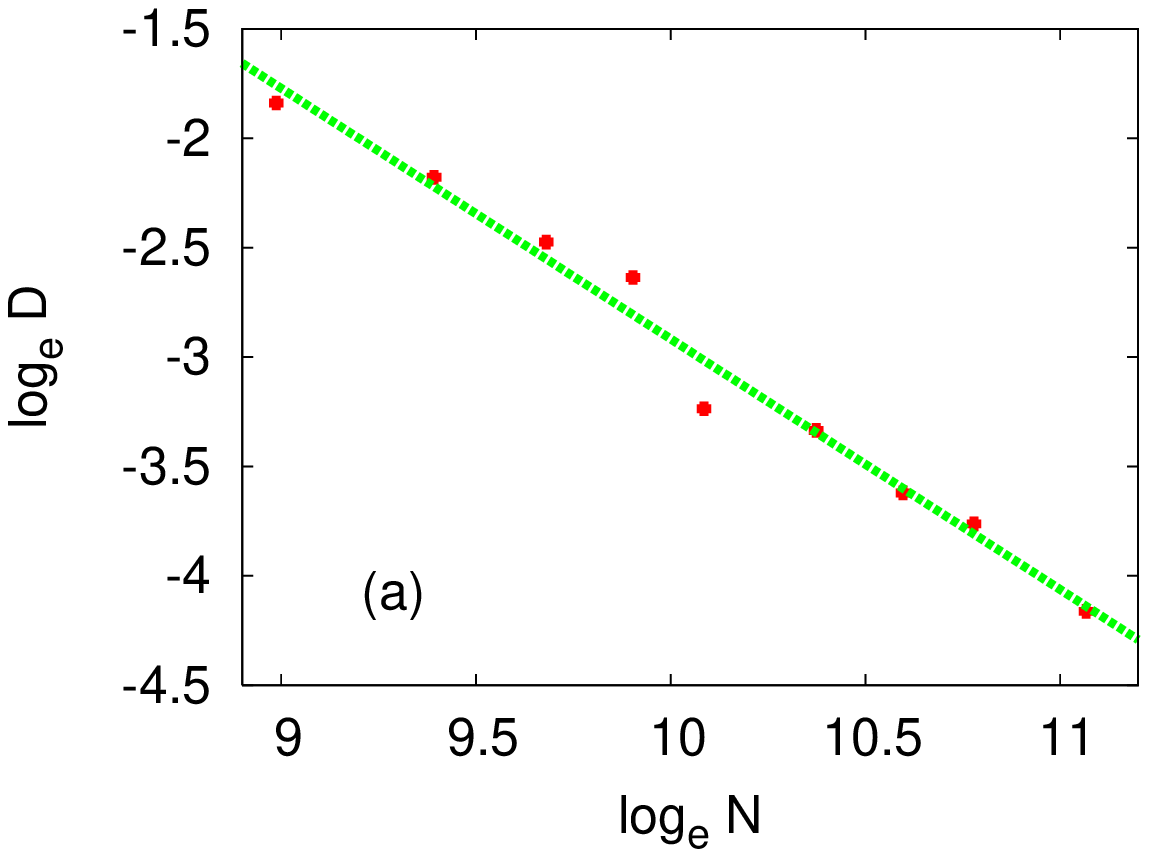}
\includegraphics[width=5cm,clip]{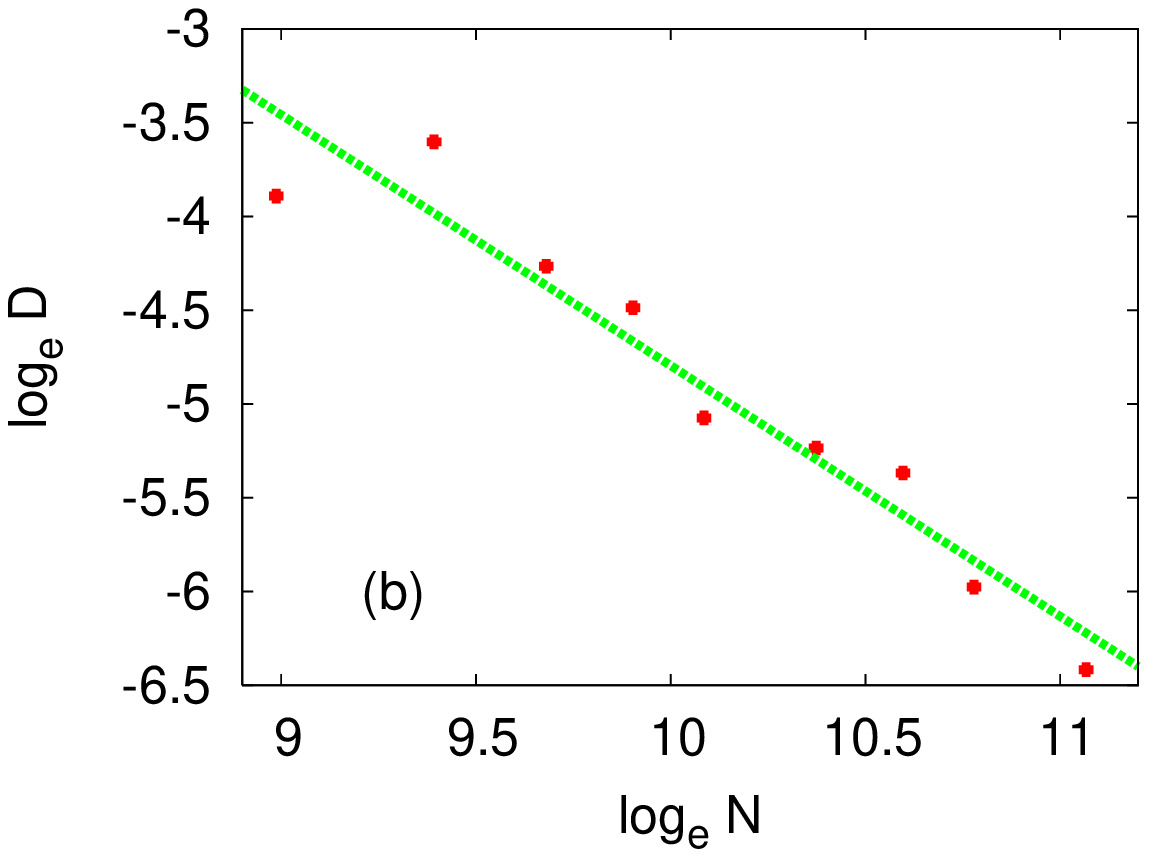}
\includegraphics[width=5cm,clip]{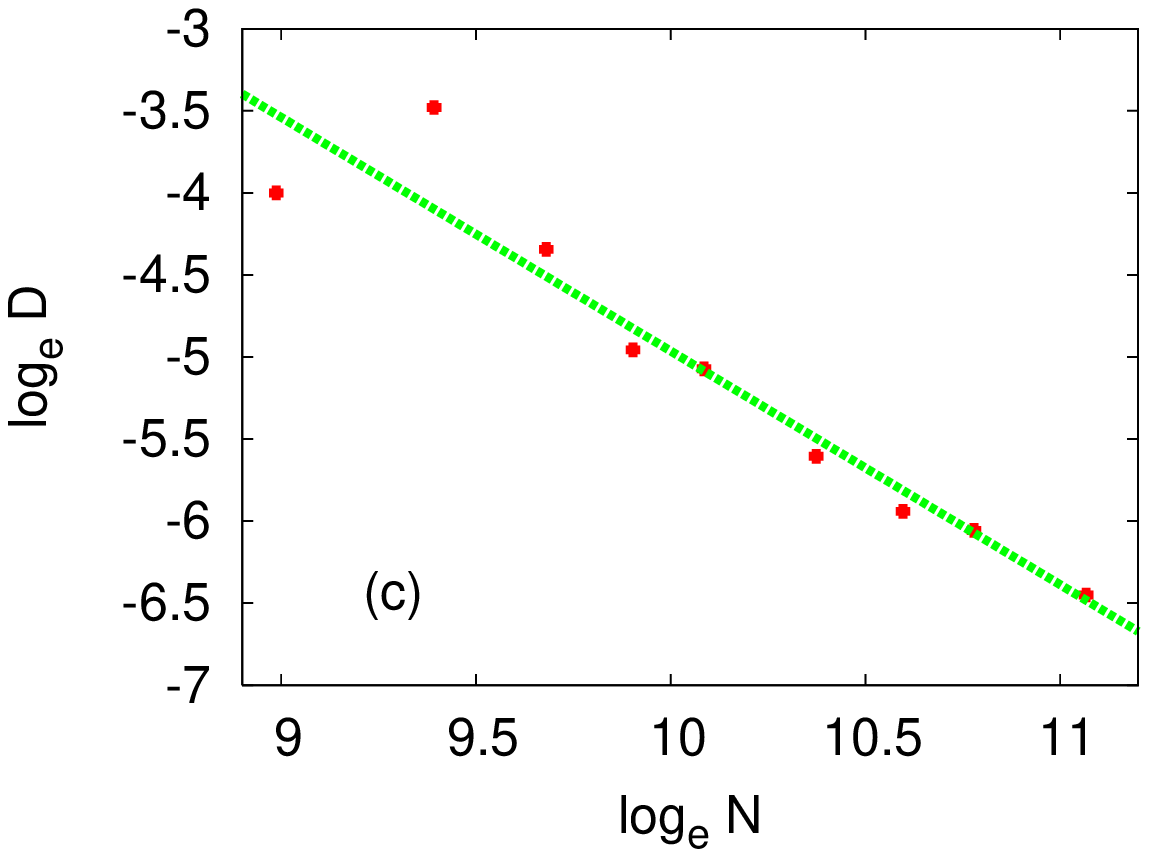}
\end{tabular}
\end{center}
\begin{center}
\begin{tabular}{lll}
\includegraphics[width=5cm,clip]{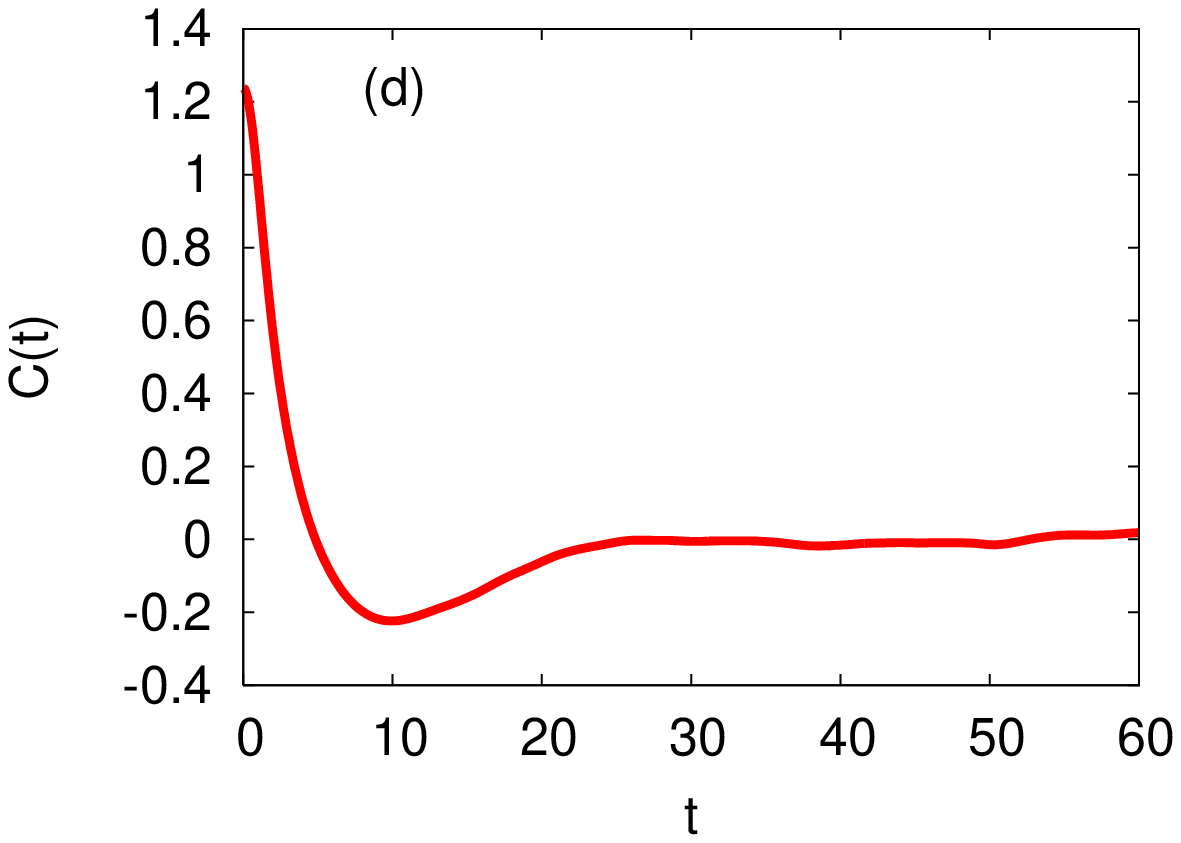}
\includegraphics[width=5cm,clip]{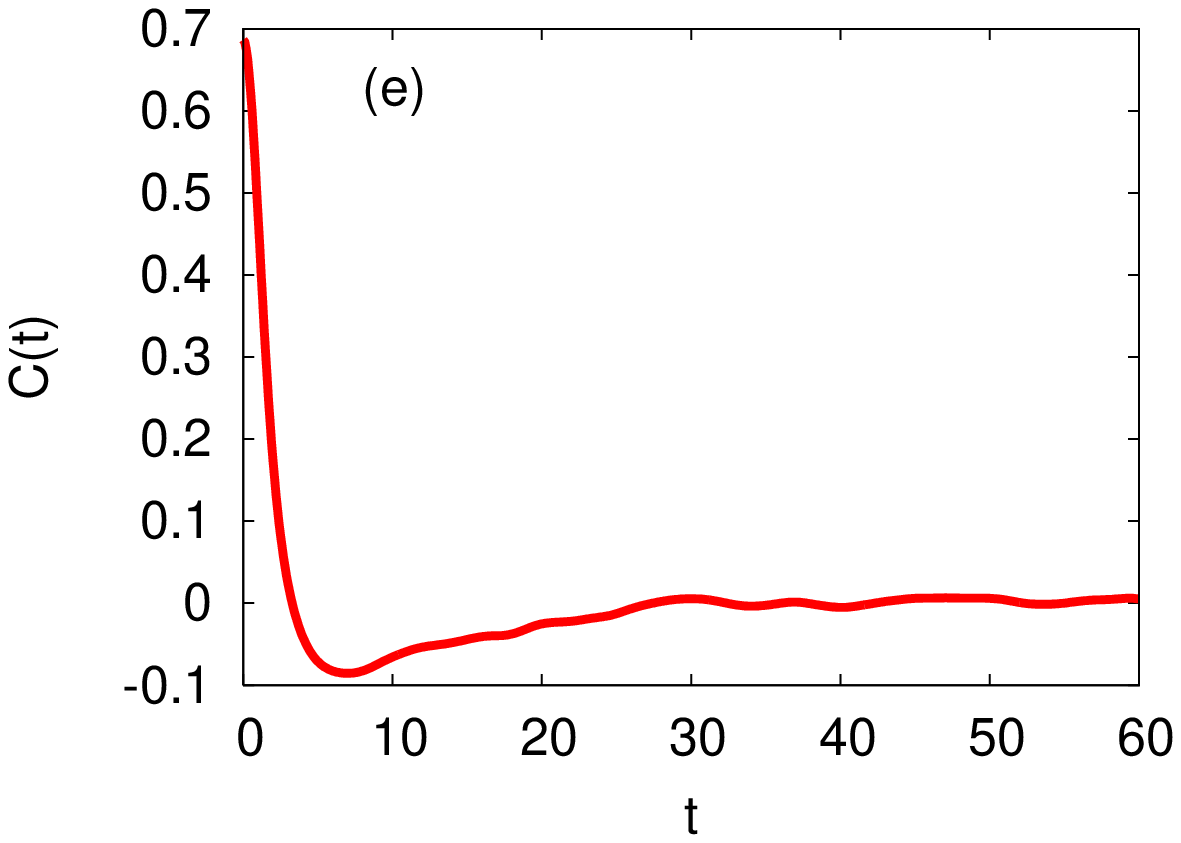}
\includegraphics[width=5cm,clip]{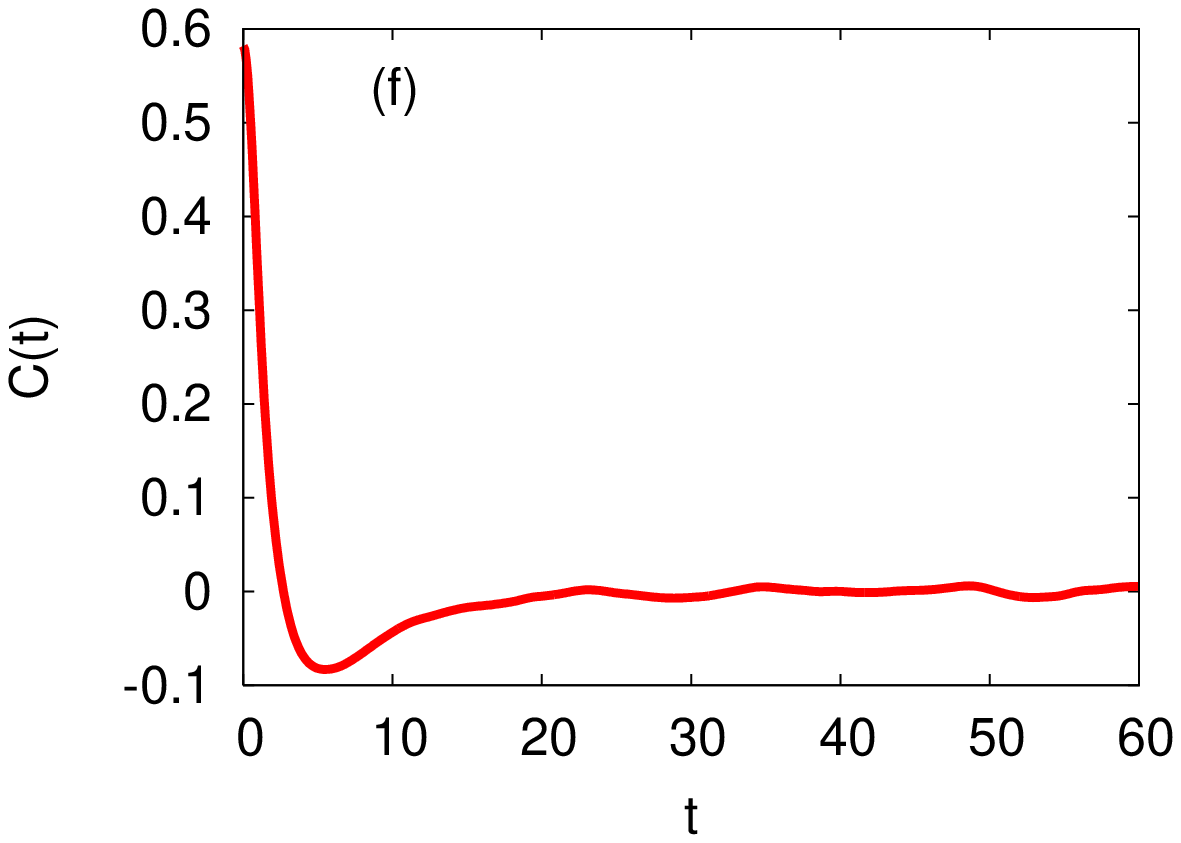}
\end{tabular}
\end{center}
\caption{
The scaling property of the diffusion coefficient $D$ with system size $N$ (upper) and the correlation function $C(t)$ (lower) in the coherent regime
of system ($\ref{phasemodel}$) with
(a)(d) $h(x) = \sin(x) - (1/2) \sin(2x)$ and $K = 1.75 > K_c = 1.59\cdots$,
(b)(e) $h(x) = \sin(x) - (1/2) \sin(3x)$ and $K = 1.675 > K_c = 1.59\cdots$,
and (c)(f) $h(x) = \sin(x+\pi /4)$ and $K = 2.04 > K_c = 1.927\cdots$,
respectively, where $N= 8000,\ldots, 64000$ for (a)-(c) and $N=24000$ for (d)-(f).
Each plot is an average over 90 samples.}
\label{fig:Dgeneral}
\end{figure}

Figures $\ref{fig:Dgeneral}$(a)-(c) show that 
$D$ decreases with $N$ as $D\sim N^{-1.15}$ for case (i), $D\sim N^{-1.34}$ for case (ii), and $D\sim N^{-1.42}$ for case (iii).
Therefore, the scaling law ($\ref{DN}$) in the coherent regime holds for these coupling functions.
Figures $\ref{fig:Dgeneral}$(d)-(f) imply $D=0$ in the limit $N \rightarrow  \infty $.
In Sec. VI, we analytically show that $D=0$ in the limit $N \rightarrow  \infty $.

\subsection{Incoherent regime}

\begin{figure}
\centering
\begin{center}
\includegraphics[width=5cm,clip]{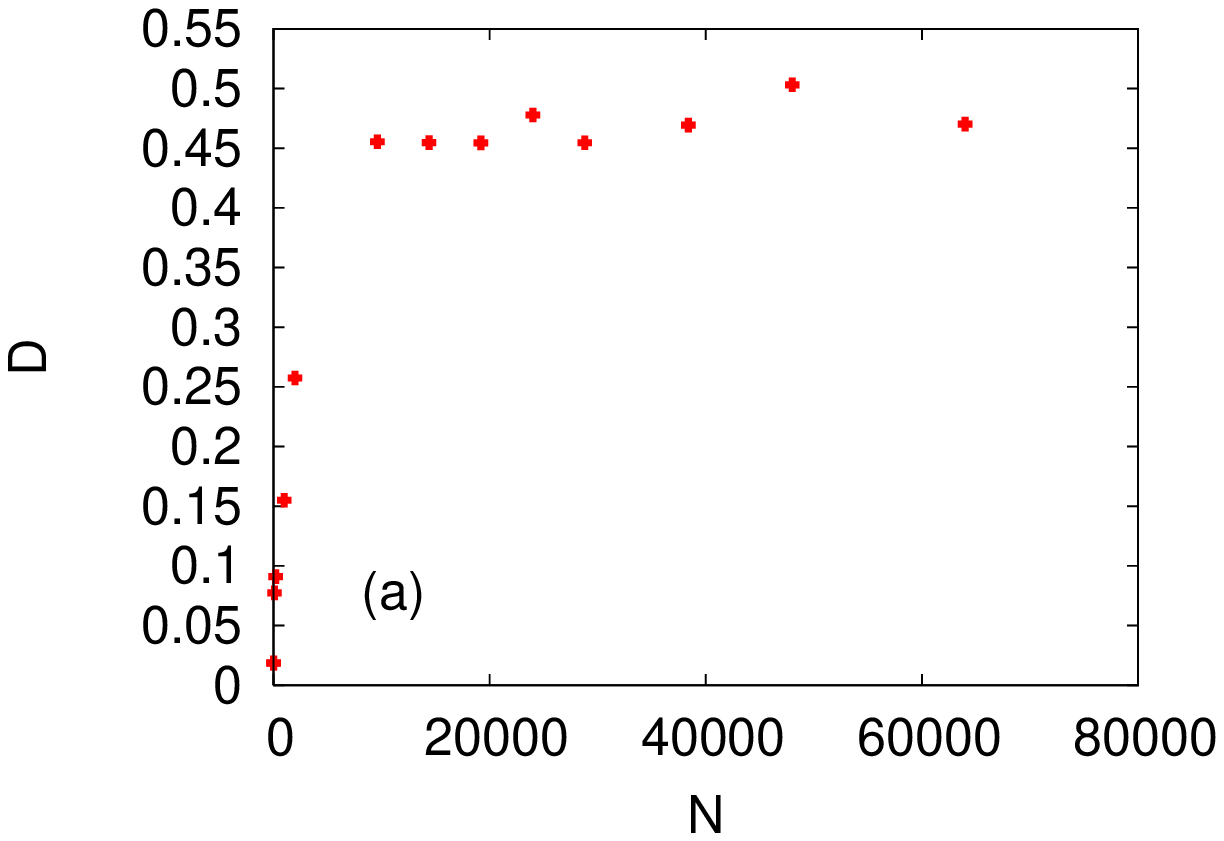}
\includegraphics[width=5cm,clip]{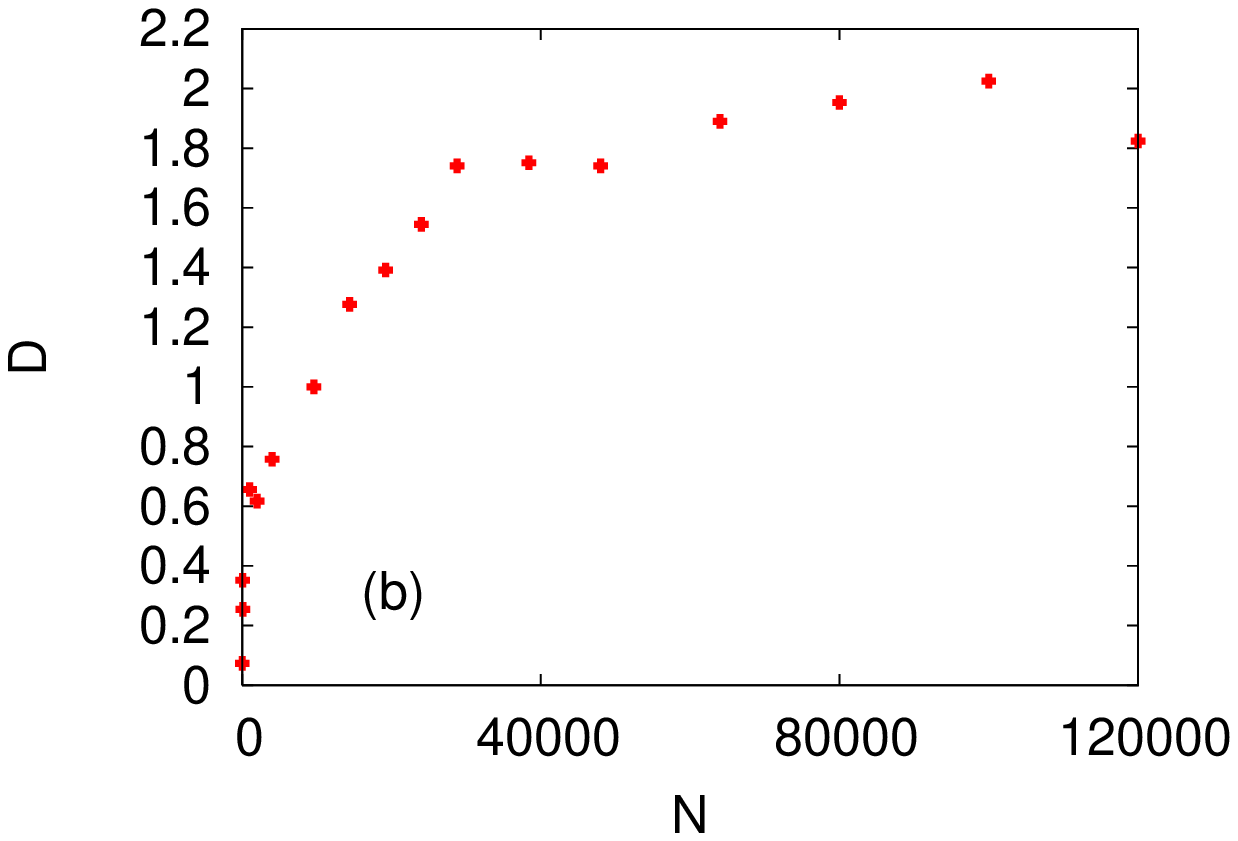}
\includegraphics[width=5cm,clip]{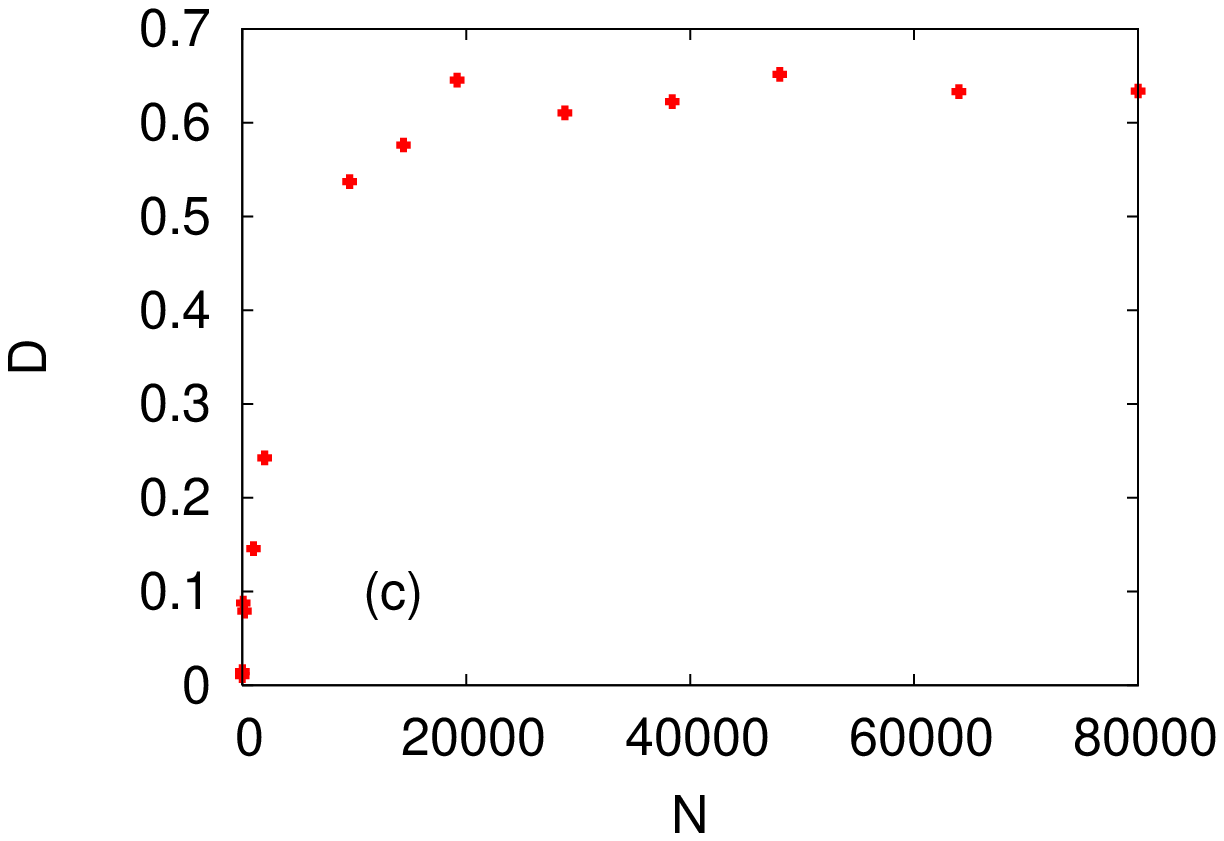}
\end{center}
\caption{The diffusion coefficient $D$ with an increase of system size $N$
in the incoherent regime of system ($\ref{phasemodel}$) with
(a) $h(x) = \sin(x+\pi /4)$ and $K = 0.8 < K_c = 1.927\cdots$, (b) $h(x) = \sin(x) - (1/2) \sin(2x)$ and $K = 1.25 < K_c = 1.59\cdots$,
and (c) $h(x) = \sin(x) - (1/2) \sin(3x)$ and $K = 0.8 < K_c = 1.59\cdots$.
$D$ fluctuates around a finite value for sufficiently large $N$.
Each plot is an average over 90 samples.
}
\label{fig:Dl3}
\end{figure}

\begin{figure}
\centering
\begin{center}
\begin{tabular}{lll}
\includegraphics[width=5cm,clip]{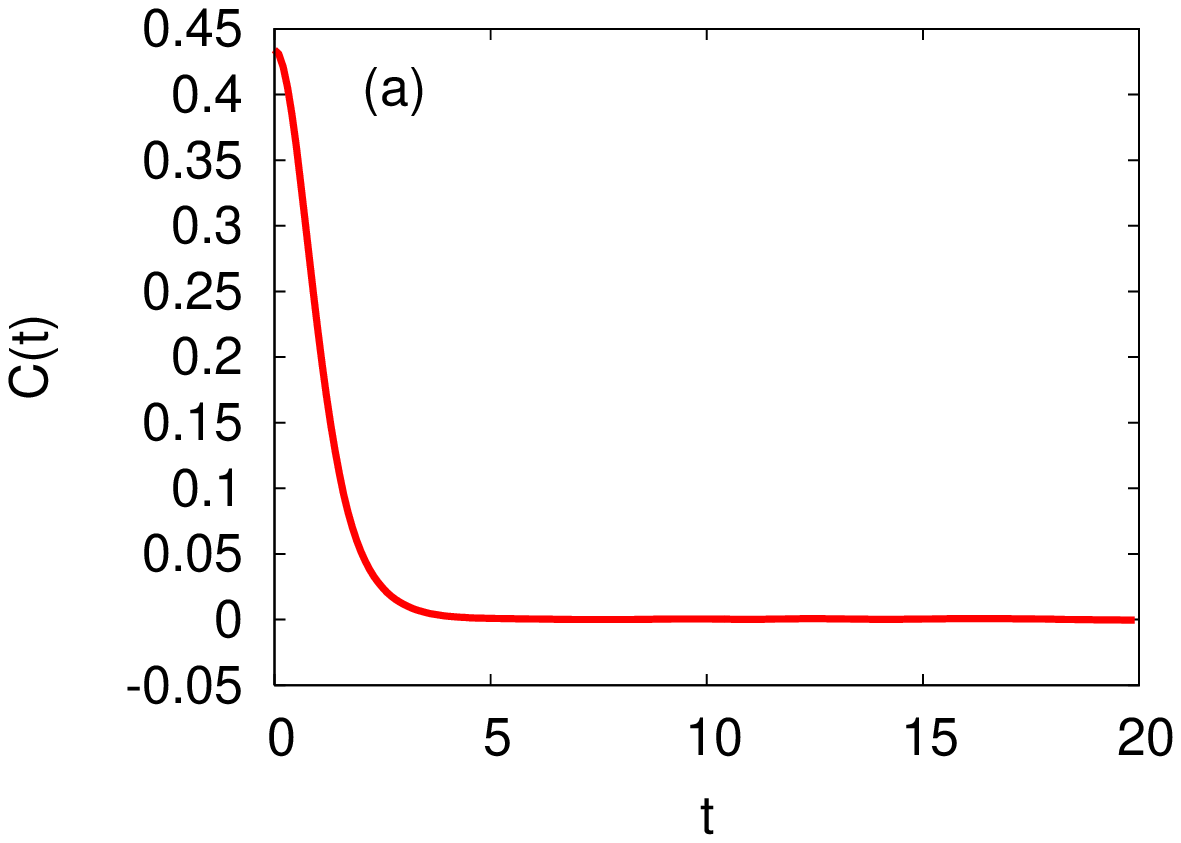}
\includegraphics[width=5cm,clip]{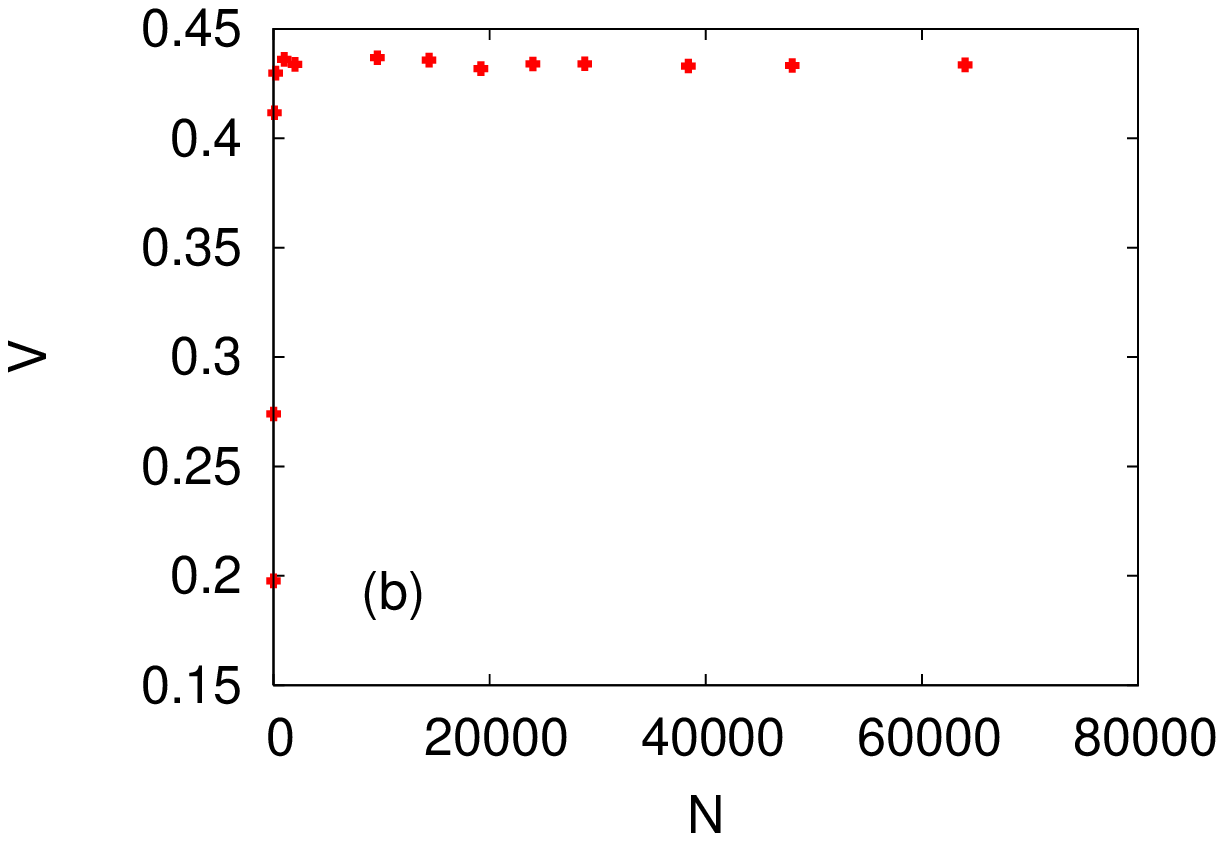}
\includegraphics[width=5cm,clip]{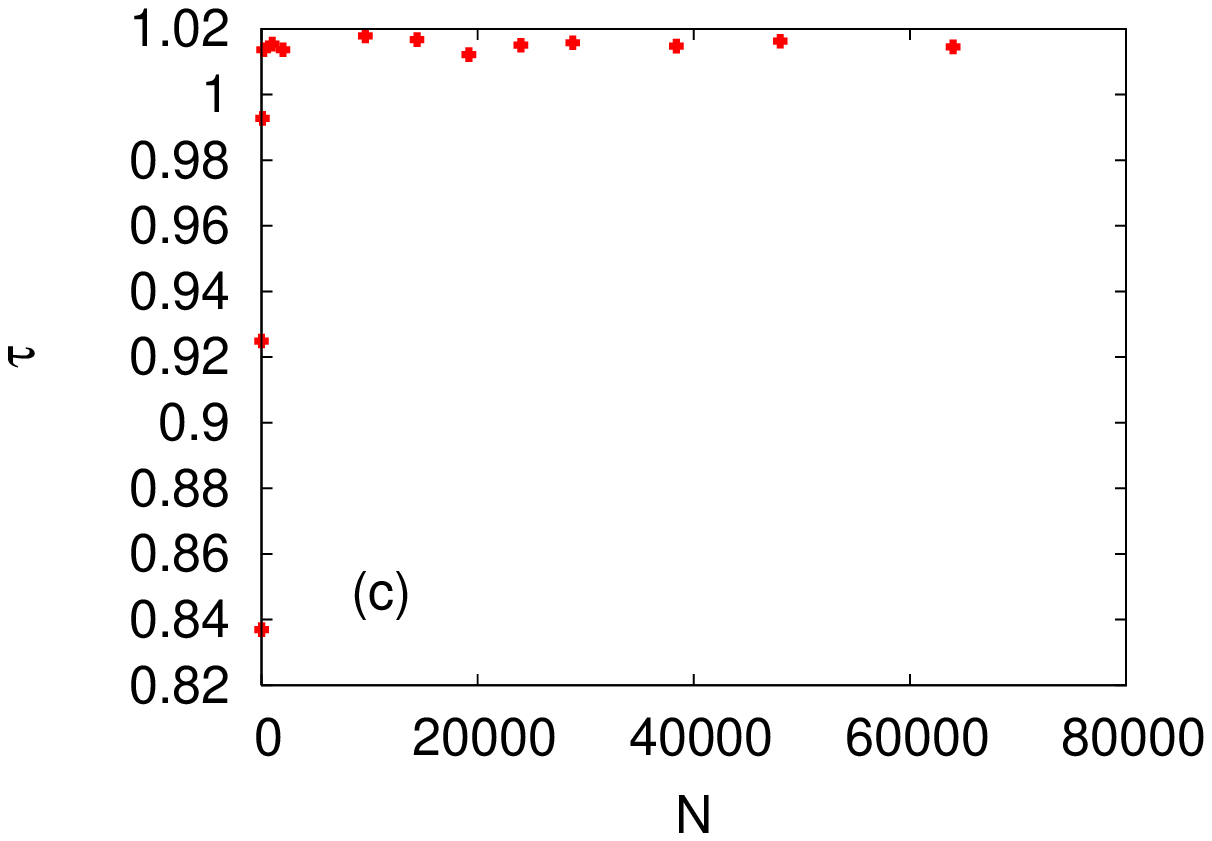}
\end{tabular}
\end{center}
\begin{center}
\begin{tabular}{lll}
\includegraphics[width=5cm,clip]{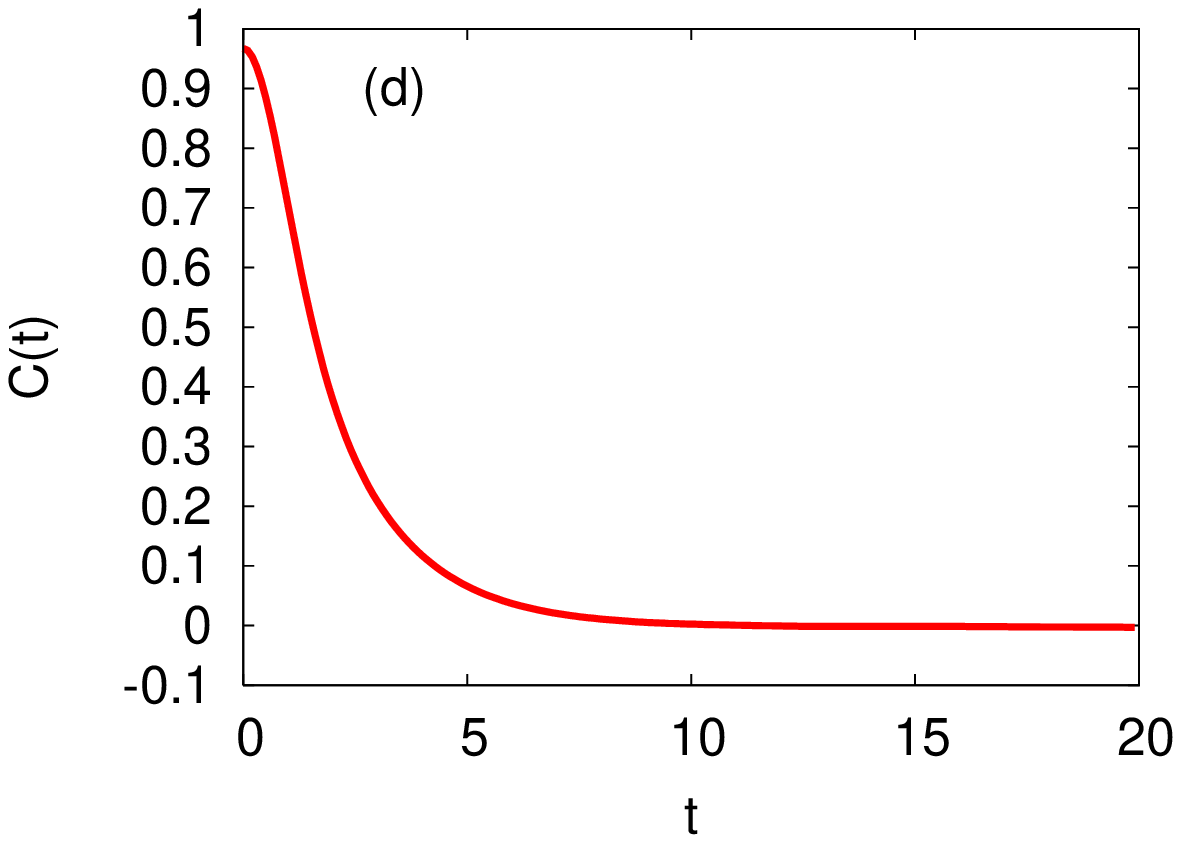}
\includegraphics[width=5cm,clip]{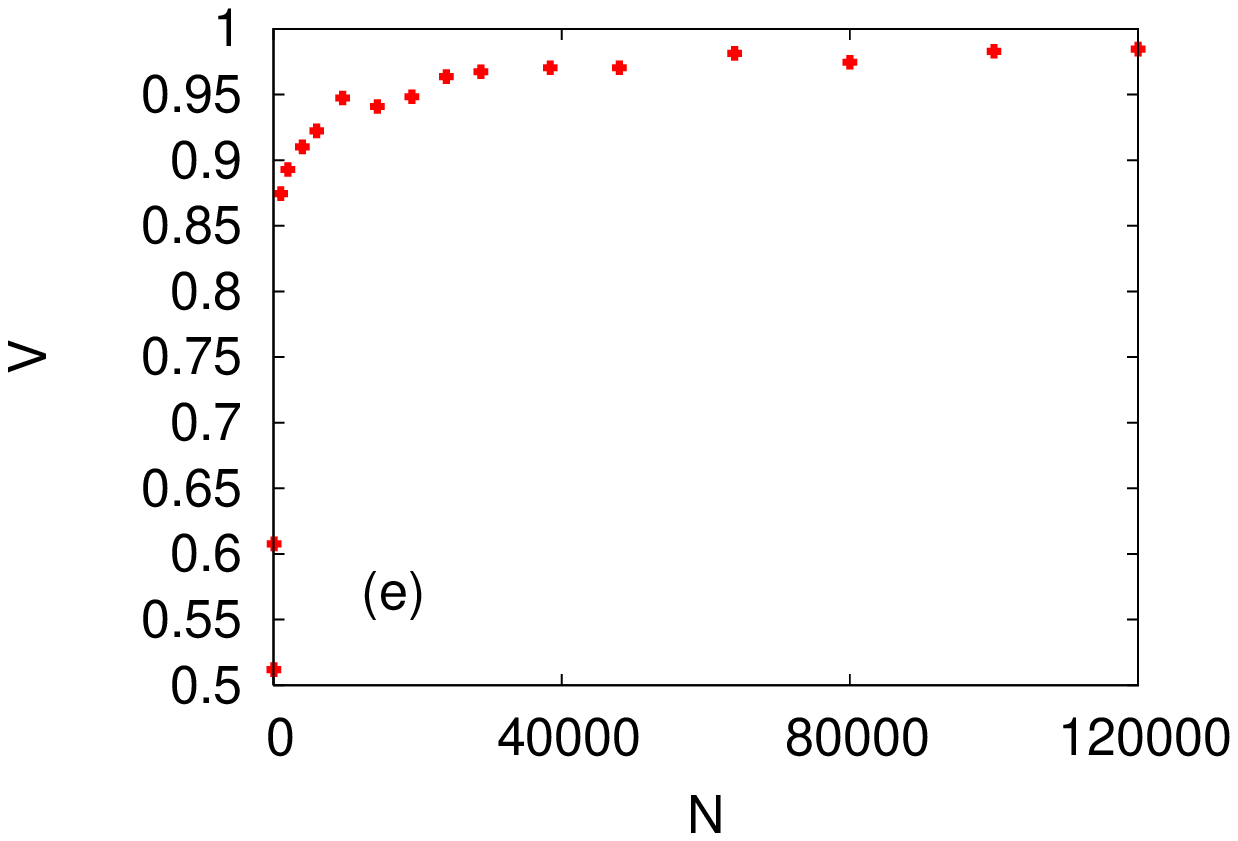}
\includegraphics[width=5cm,clip]{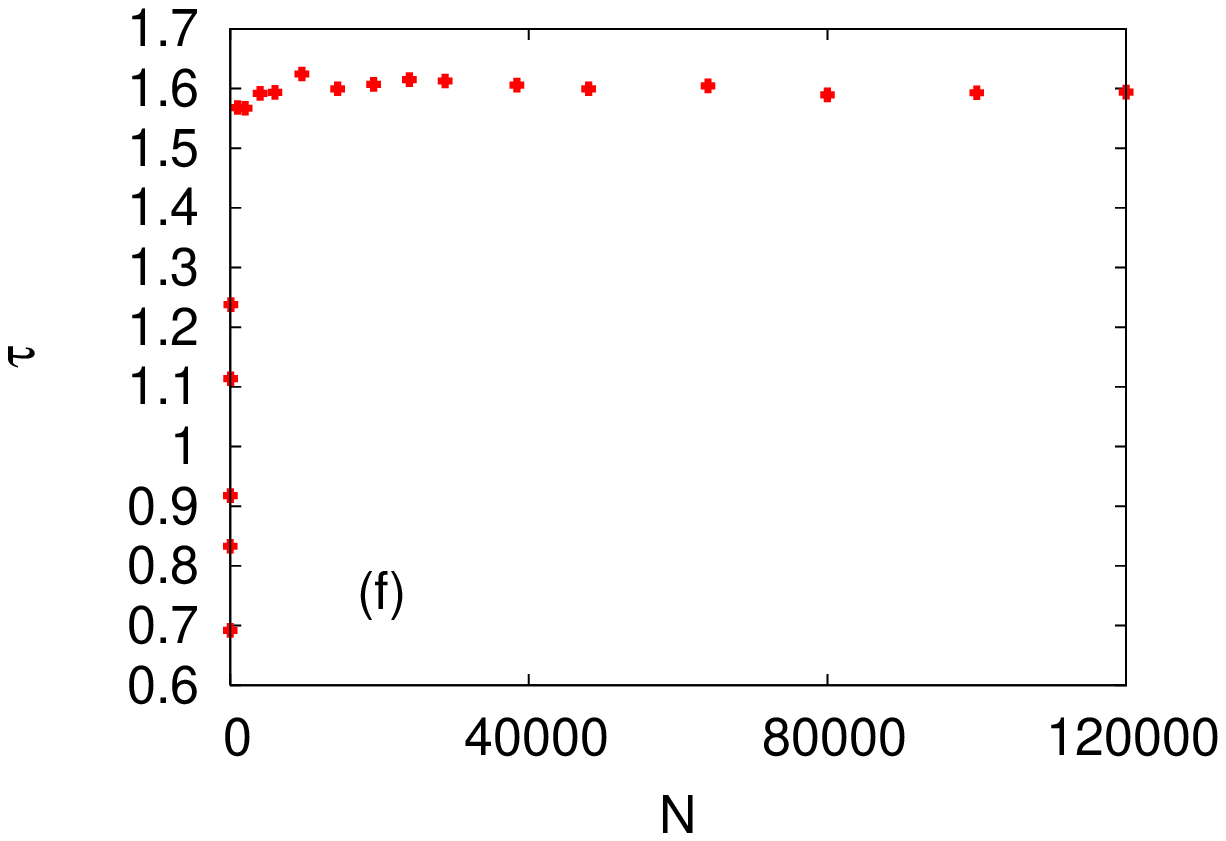}
\end{tabular}
\end{center}
\begin{center}
\begin{tabular}{lll}
\includegraphics[width=5cm,clip]{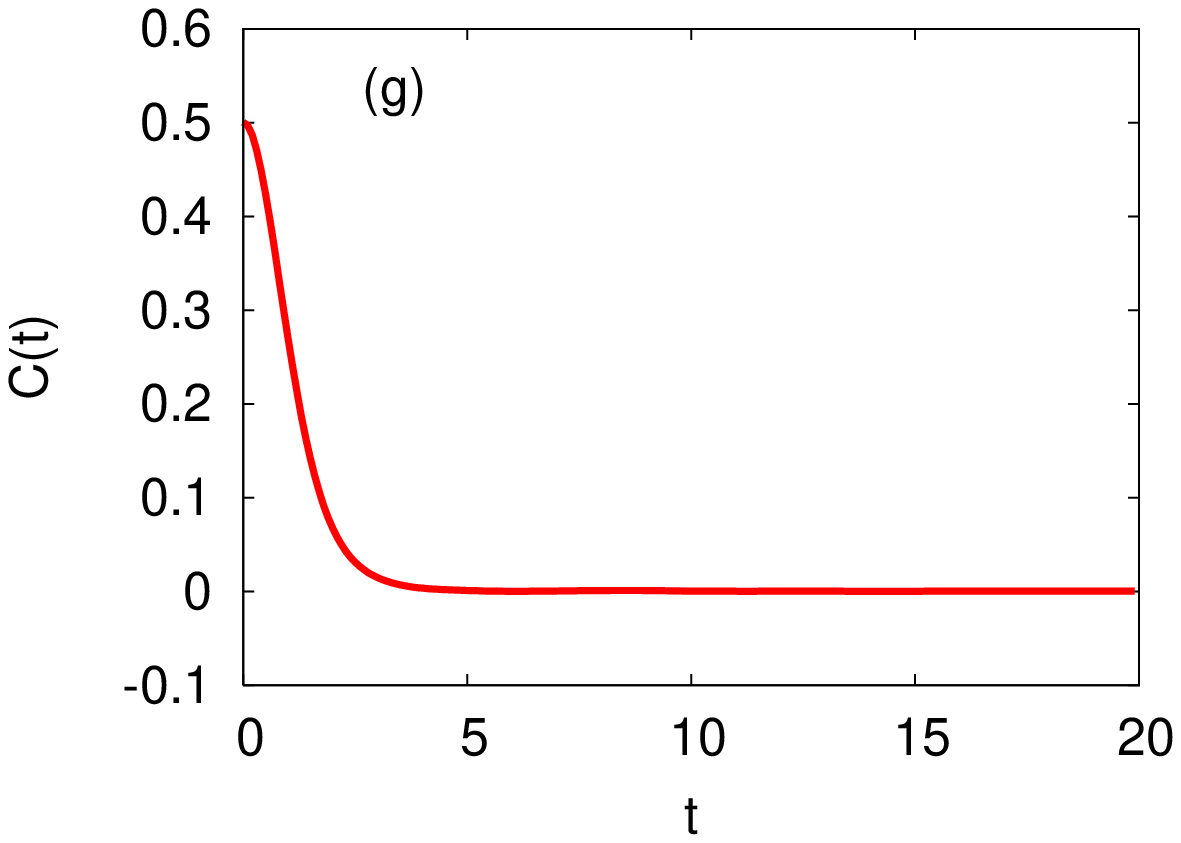}
\includegraphics[width=5cm,clip]{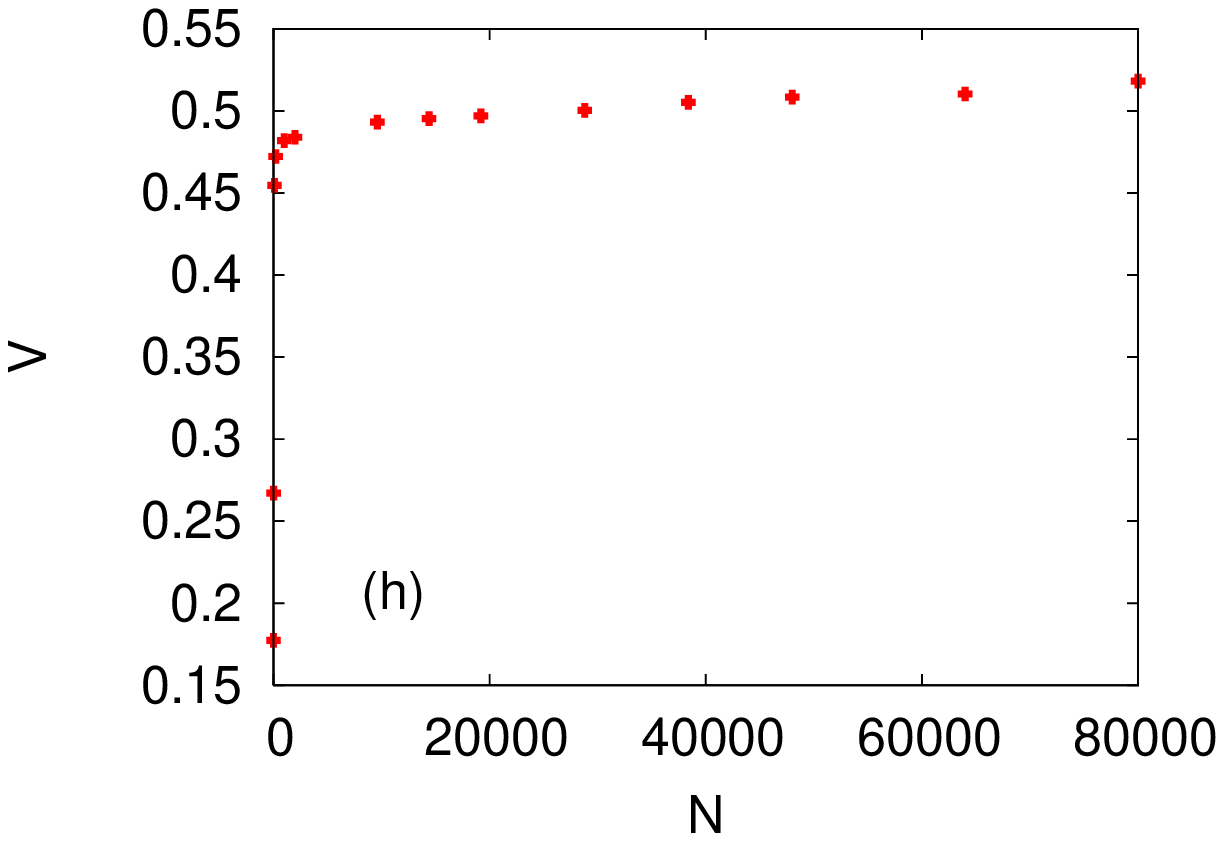}
\includegraphics[width=5cm,clip]{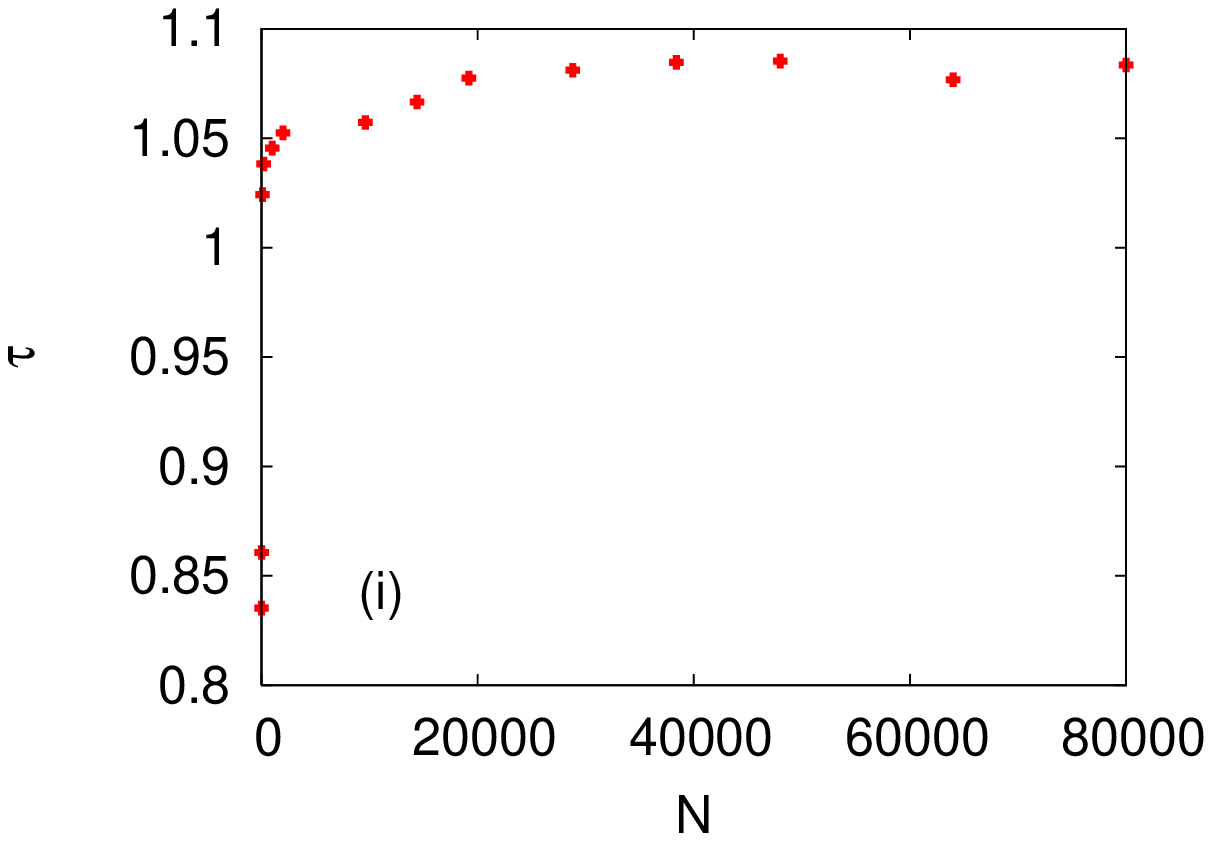}
\end{tabular}
\end{center}
\caption{The correlation function $C(t)$ for $N=64000$ (left), the variance $V$ (middle), and the correlation time $\tau $ (right) of the order parameter
in the incoherent regime of system ($\ref{phasemodel}$) with
(a)(b)(c) $h(x) = \sin(x+\pi /4)$ and $K = 0.8 < K_c = 1.927\cdots$, (d)(e)(f) $h(x) = \sin(x) - (1/2) \sin(2x)$ and $K = 1.25 < K_c = 1.59\cdots$,
and (g)(h)(i) $h(x) = \sin(x) - (1/2) \sin(3x)$ and $K = 0.8 < K_c = 1.59\cdots$.
$C(t)$ almost exponentially decreases with $t$.
$V$ and $\tau $ fluctuate around a finite value for sufficiently large $N$.
Each plot is an average over 90 samples.}
\label{fig:Cl3}
\end{figure}

We demonstrate that $D\sim O(1)$ for any large $N$ in cases (i)-(iii).
As shown in Figs.~$\ref{fig:Dl3}$(a)-(c),
$D$ fluctuates around a finite value for sufficiently large $N$.
It implies $D\sim O(1)$.
Further, we support this scaling as follows.
The forms of the correlation function of the order parameter are almost exponential as shown in Figs.~$\ref{fig:Cl3}$(a), $\ref{fig:Cl3}$(d), and $\ref{fig:Cl3}$(g).
In addition, our numerical simulations show that
$V \sim O(1)$ \cite{Daido,Hildebrand,Buice} and $\tau \sim O(1)$ \cite{Goldenfeldbook,Nishimori}
as shown in Figs.~$\ref{fig:Cl3}$(b)(c), $\ref{fig:Cl3}$(e)(f), and $\ref{fig:Cl3}$(h)(i).
Therefore, we conclude $D\sim O(1)$ from the same argument in Sec. IIIB.

\section{Derivation of $D=0$ for the Kuramoto model}

In this section, we analytically demonstrate that $D=0$ in the limit $N\rightarrow  \infty $ in the coherent regime of the Kuramoto model
by reference to Daido \cite{Daido}, which deals with statistical properties in the vicinity of the transition point.
In general, fluctuations are amplified near the transition point \cite{Goldenfeldbook,Nishimori}.
Therefore, if $D=0$ is derived in the limit $K\rightarrow  K_c +0$, then $D=0$ would be justified for $K>K_c$.

We define the complex order parameter \cite{Daido} as follows:
\begin{align}
\label{Complex}
Z(t) \equiv \frac {1}{N} \sum _{j=1} ^N \exp (2\pi i(\theta _j - \Omega t)),
\end{align}
where $\Omega $ represents the frequency of entrainment and $|Z(t)| = R(t)$.
We denote the diffusion coefficient of $\int _0 ^t Z(s) ds$ by $D_Z$,
which is defined as follows:
\begin{align}
\label{DZ}
D_Z \equiv \lim _{t\rightarrow \infty } \sigma^2 _Z (t) /  2t,
\end{align}
where the variance of $\int _0 ^t Z(s) ds$ is given by
\begin{align}
\label{sigmaZ}
\sigma^2 _Z(t) \equiv N  \left[ \left\langle \left| \int _{t_0} ^{t+t_0} Z(s)ds - \langle Z \rangle _t  t \right|^2  \right\rangle _t \right] _s.
\end{align}
With the correlation function of $Z(t)$ given by
\begin{align}
\label{CZ}
C_Z(t) \equiv N  [ \langle ( Z(t+t_0) - \langle Z \rangle _t)( Z^{*}(t_0) - \langle Z^{*} \rangle _t) \rangle _t  ]_s,
\end{align}
where $Z^{*}$ represents the complex conjugate of $Z$,
$D_Z$ satisfies the following equation \cite{Correlationfunction}:
\begin{align}
\label{CDZ}
D_Z = \frac{1}{2} \int _ {-\infty} ^{\infty} C_Z(s) ds.
\end{align}
In the limit $N\rightarrow \infty $,
$C_Z (t)$
takes the following form \cite{Daido}:
\begin{align}
C_Z (t) \sim \Pi (\sqrt{Q} t) / \sqrt{Q},
\end{align}
where $\Pi (x) = e^{-s|x|} + e^{-\sqrt{5} s|x|}/\sqrt{5} - s|x|(e^{-s|x|}+e^{-\sqrt{5}s|x|})$ with a certain constant $s$ and $Q = Q(K)$.
Integrating the above equation by $t$ for $t\in (-\infty , \infty )$, we can obtain $D_Z=0$ from Eq.~($\ref{CDZ}$).

Next, let us prove that in the limit of $N \rightarrow \infty $, if $D_Z=0$ then $D=0$.
We introduce a variable $w(t)$ for representing the fluctuations of $Z(t)$ as follows:
\begin{align}
\label{fluctuation_pre}
w(t) \equiv Z (t) - \hat {Z},
\end{align}
where
\begin{align}
\label{infinity-orderparameter_pre}
\hat{Z} \equiv  \langle \lim _{N \rightarrow \infty} Z \rangle _t,
\end{align}
and $\hat {Z}  =  \mathrm{const.} \not = 0$ in the coherent regime (see p.29 in \cite{Crawford2}).
For large $N$, $w(t)$ is small \cite{Daido} in the regime after an initial transient period \cite{assumption_pre}.
Hereafter, let us assume that $t=0$ is included in that regime.
The deviation $w(t)$ is scaled as $ O(1/\sqrt{N})$ \cite{Daido,assumption_pre}.
It should be noted that we can assume $\hat{Z}$ to be a positive real number because of the rotation symmetry of the system \cite{assumption_pre}.
Then, $R(t)$ can be approximated as follows:
\begin{align}\label{proof}
\nonumber
R(t)  &=\sqrt{(\hat{Z} + \mathrm{Re}(w(t)))^2 + (\mathrm{Im}(w(t)))^2}
\\\nonumber
&\approx \hat{Z} \sqrt{1+2\mathrm{Re}(w(t)) / \hat{Z} }
\\
&\approx \hat{Z} + \mathrm{Re}(w(t)),
\end{align}
where we have neglected the higher-order terms that vanish in the limit $N\rightarrow \infty$.
From Eq.~($\ref{proof}$),
if the diffusion coefficient of $\int _0 ^t \mathrm{Re}(w(s)) ds$ is equal to 0,
then that of $\int _0 ^t R(s) ds$ is also equal to 0 because $\hat{Z}$ is constant.
That is,
in the limit $N\rightarrow \infty $,
if $D_Z=0$ then $D=0$.
Hence, because $D_Z=0$ in the limit $N\rightarrow \infty $ as explained in the previous paragraph, $D=0$ in the same limit.

\section{Derivation of $D=0$ for a general coupling function}

In this section, we analytically demonstrate that $D=0$ in the limit $N\rightarrow  \infty $
in the coherent regime of the phase oscillator model $(\ref{phasemodel})$ with a general coupling function $h(x)$
by reference to Daido \cite{Daido}, which considered the sinusoidal coupling function. 

\begin{figure}
\centering
\begin{tabular}{ll}
\includegraphics[width=3.5cm,clip]{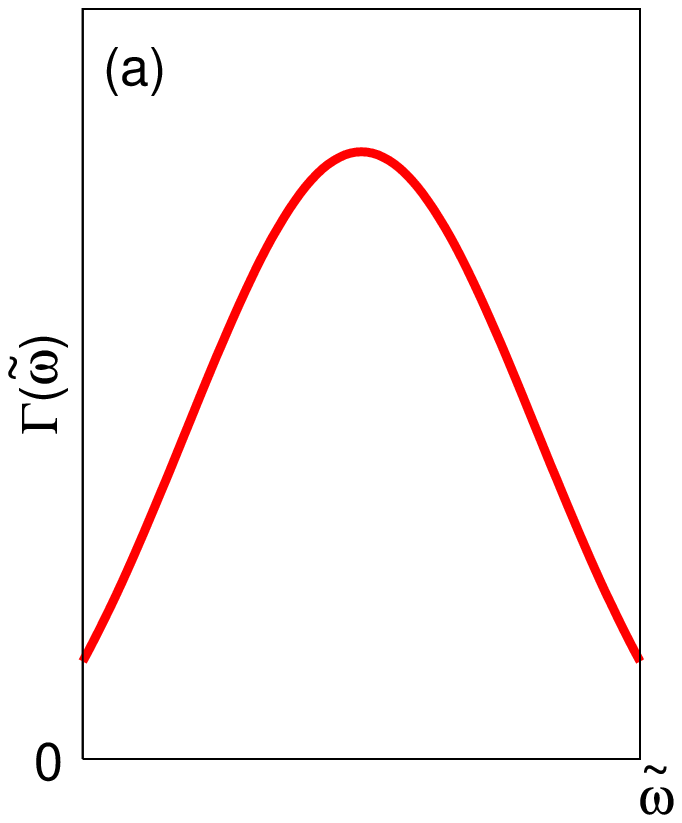}
\includegraphics[width=3.5cm,clip]{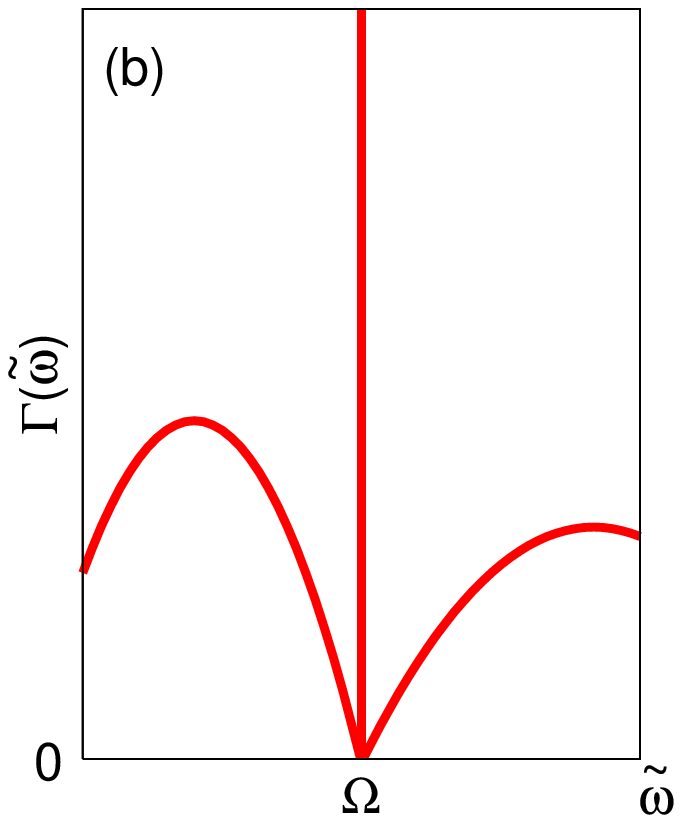}
\end{tabular}
\caption{\label{fig:schematic2}
Schematic view of $\Gamma (\tilde \omega )$. (a) An incoherent state. (b) A coherent state.
}
\end{figure}

Let us explain a key assumption for deriving $D=0$.
When the system shows synchronization,
the oscillators are divided into the following two groups:
(i) entrained oscillators, which are synchronized with the frequency $\Omega$,
and (ii) nonentrained oscillators, which are not synchronized.
Entrained oscillators play a role in reducing the fluctuations of the order parameter \cite{Daido4}.
On the other hand, nonentrained oscillators show minor fluctuations \cite{Daido4}.
We denote
by $\Gamma (\tilde \omega )$ the distribution of the mean frequencies $\tilde \omega $
of the oscillators.
Here, we assume that the following condition holds in the coherent regime:
\begin{align}
\label{criticaleq}
\lim _{\tilde \omega \rightarrow \Omega } \Gamma (\tilde \omega ) = 0  \ \mathrm{with} \ N \rightarrow \infty,
\end{align}
where
$\Omega $ is the common frequency of the entrained oscillators.
Daido \cite{Daido3} analytically demonstrated that the above condition holds
if the coupling function $h(x)$ exhibits only one local minimum and only one local maximum in its domain.
Further, Daido \cite{Daido2} numerically confirmed that this condition holds for more general coupling functions.
Figure $\ref{fig:schematic2}$ illustrates a typical example,
where Eq.~$(\ref{criticaleq})$ holds in a coherent state whereas it does not hold in an incoherent state \cite{PikovskyBook}.
Equation $(\ref{criticaleq})$ means that
the density of non-entrained oscillators with mean frequencies close to but not equal to $\Omega $ significantly
decreases as $\tilde \omega  \rightarrow  \Omega $.
As a result, fluctuations become minor \cite{Daido4}
because the more distant the mean frequency of an oscillator is from $\Omega $,
the smaller are the fluctuations caused by the oscillator.

Considering the power spectrum of $\sqrt{N}  (R(t) - \langle R \rangle _t)$,
its asymptotic form in the limit $N\rightarrow \infty$ is given by $I(\omega ) = \int _{-\infty } ^{\infty } C(s) e^{i\omega s} ds$ \cite{relation}.
From Eq.~($\ref{CD}$), we obtain
\begin{align}
\label{inotherwords2}
D= \frac{1}{2} \lim _{\omega \rightarrow 0}I(\omega).
\end{align}
This equation indicates that
the value of $D$ is almost determined by the value of $\Gamma (\tilde \omega )$ around $\tilde \omega = \Omega $,
since $\Omega $ can be replaced by 0 due to a transformation of the phase variables.
Therefore, it is natural that the value of $D$ becomes very small if Eq.~$(\ref{criticaleq})$ holds.

This section is organized as follows.
First, we transform the original equation ($\ref{phasemodel}$) into another expression according to Ref. \cite{Daido3}.
Next, we derive a self-consistent equation governing the fluctuations of the order parameter.
Finally, we show $D=0$ by using the Fourier transform of the self-consistent equation.

\subsection{Transformation of the phase oscillator model}

The coupling function $h$ in Eq.~($\ref{phasemodel}$) can be generally represented by the Fourier series as follows:
\begin{align}
\label{Fourierseries}
h(x) = \sum _l q_l e^{2\pi i l x},
\end{align}
where $q_l$ represents the $l$th Fourier coefficient for $l=\pm 1,\pm 2,\cdots$.
We assume that all the synchronized oscillators rotate with the common frequency $\Omega $ \cite{Daido3}
and introduce the generalized complex order parameters as follows:
\begin{align}
\label{gComplex}
Z_l(t) \equiv \frac {1}{N} \sum _{j=1} ^N \exp (2\pi i l(\theta _j - \Omega t)),
\end{align}
where $Z_1 (t) = Z(t)$.
By using $q_l$ and $Z_l$, the order function $H(x)$ \cite{Daido3} is defined as follows:
\begin{align}
\label{orderfunction}
H(x) \equiv - \sum _{l} q_l Z_l e^{- 2\pi i l x}.
\end{align}
From Eq.~$(\ref{orderfunction})$, we can transform Eq.~($\ref{phasemodel}$) into the following form:
\begin{align}
\label{phasemodel2}
\dot \theta_j = \omega_j - K H(\theta _j - \Omega t).
\end{align}
By introducing new variables $\tilde \theta _j \equiv \theta _j - \Omega t$ and $\Delta _j \equiv \omega_j - \Omega $,
Eq.~($\ref{phasemodel2}$) is transformed into
\begin{align}
\label{phasemodel3}
d \tilde \theta _j / dt = \Delta _j - K H(\tilde \theta _j).
\end{align}

\subsection{The self-consistent equation of fluctuations}

First, let us introduce a new variable $w_l$ for representing the fluctuations of $Z_l(t)$ as follows:
\begin{align}
\label{fluctuation}
w_l(t) \equiv Z_l (t) - \hat Z_l,
\end{align}
where
\begin{align}
\label{infinity-orderparameter}
\hat{Z_l} \equiv \langle \lim _{N \rightarrow \infty} Z_l \rangle _t,
\end{align}
and $\hat {Z_l}  =  \mathrm{const.} \not = 0$ in the coherent regime (see p.29 in \cite{Crawford2}).
For large $N$, $w_l(t)$ is small \cite{Daido} in the regime after an initial transient period \cite{assumption}.
Hereafter, let us suppose that $t=0$ is included in that regime.

Now we assume that $\tilde \theta _j$ can be divided into two parts as follows \cite{Daido},
\begin{align}
\label{twoparts}
\tilde \theta_ j = \psi _j + \phi _j,
\end{align}
where $\psi _j$ and $\phi _j$ correspond to the dominant phase motion and the small deviation from it, respectively.
The dominant phase motion can be described as follows:
\begin{align}
\label{dominant}
d \psi _j / dt &\equiv \Delta _j - K \hat H(\psi _j),
\\\nonumber
\psi _j (0) &= \tilde \theta _j (0) = \theta _j (0),
\end{align}
where
\begin{align}
\label{orderfunction2}
\hat H(x) \equiv - \sum _{l} q_l \hat Z_l e^{- 2\pi i l x}.
\end{align}
That is, $\psi _j$ corresponds to $\tilde \theta_ j$ in the infinite-size system.

Next, let us introduce a self-consistent equation of $w_l(t)$.
If we put
\begin{align}
\label{normalized1}
\tilde w_l \equiv \sqrt{N}  w_l,
\end{align}
then $\tilde {w_{l}}$ is $O(1)$, as discussed in p.760 of \cite{Daido}.
The deviation $\phi _j$ induced by $w_l$ should be of $O(N^{-1/2})$,
so that we can expand $\tilde \theta _j$ in $N^{-1/2}$ as follows:
\begin{align}
\label{normalized2}
\tilde \theta _j &= \psi _j + \phi _j,
\\  &= \psi _j + \frac{\tilde \phi _j}{{\sqrt{N}}} + O(N^{-1}).\nonumber
\end{align}
Substituting Eqs.~$(\ref{fluctuation})$, $(\ref{normalized1})$, and $(\ref{normalized2})$ into Eq.~$(\ref{phasemodel3})$
and comparing $O(N^{-1/2})$ terms, we obtain
\begin{align}
\label{expand1}
d\tilde \phi _j / dt = K \sum _l q_l (-2\pi i l \hat Z_l \tilde \phi _j + \tilde w_l) e^{-2\pi i l\psi _j}.
\end{align}
Furthermore, from Eqs.~$(\ref{gComplex})$, $(\ref{fluctuation})$, $(\ref{normalized1})$, and $(\ref{normalized2})$,
we can derive
\begin{align}
\label{expand2}
\nonumber \tilde w_l = &\sqrt{N} ( - \hat Z_l + N^{-1} \sum _{j=1} ^{N} e^{2\pi i l\psi _j})
\\&+ 2\pi i l N^{-1} \sum _{j=1} ^{N} \tilde \phi _j e^{2\pi i l\psi _j} + O(N^{-1/2}).
\end{align}
Note that the first term on the right-hand side (r.h.s.) of Eq.~$(\ref{expand2})$ is $O(1)$ \cite{Daido}.
By inserting the solutions of Eq.~$(\ref{expand1})$ into Eq.~$(\ref{expand2})$ and by considering the limit $N\rightarrow \infty $,
we arrive at the self-consistent equations for $\tilde w_l$ as follows:
\begin{align}
\label{self-consistent1}
\displaystyle \tilde w_l (t) = P_l(t)+ 2\pi i l K \sum _{l'} q_{l'} \int _0 ^t dt'  A_{l, l'}(t,t') \tilde w_{l'}(t'),
\end{align}
where
\begin{align}
\label{self-consistent3}
P_l (t) \equiv \lim _{N\rightarrow \infty } \sqrt{N} ( - \hat Z_l + N^{-1} \sum _{j=1} ^{N} e^{2\pi i l\psi _j}),
\end{align}
and the kernel $A_{l,l'}$ is defined by
\begin{align}
\label{self-consistent2}
\nonumber \displaystyle A_{l, l'}(t,t') \equiv &\lim _{N\rightarrow \infty } N^{-1} \sum _{j=1} ^{N} \exp \Bigl\{ 2\pi i (l \psi _j(t) - l' \psi _j (t'))
\\&- K \int _{t'} ^{t} d\tau \hat H'(\psi _j (\tau ))\Bigr\}.
\end{align}
Here $\hat H'(x)$ represents the derivative of $\hat H(x)$ with respect to $x$.

\subsection{The Fourier transform of the self-consistent equation}

The goal of this subsection is to show that $D=0$ in the limit $N \rightarrow \infty $ under assumption $(\ref{criticaleq})$.
From the discussion in Sec. V, $D=0$ if $D_Z = 0$.
Further, the condition $D_Z = 0$ is equivalent to $\lim _{\omega \rightarrow 0}I_Z(\omega) =0$
where $I_Z(\omega ) = \int _{-\infty } ^{\infty } C_Z(s) e^{i\omega s} ds$
is the asymptotic form of the power spectrum of $\sqrt{N}  (Z(t) - \langle Z \rangle _t)$ in the limit $N\rightarrow \infty$.
Therefore, it is only necessary to show $\lim _{\omega \rightarrow 0}I_Z(\omega) =0$.
To evaluate $I_Z (\omega )$,
we cast $\tilde w_l$ and $P_l(t)$ into the form of the frequency domain representation, respectively, as follows \cite{Fourier}:
\begin{align}
\label{Fourier}
\tilde w_l ^* (\omega ) \equiv  \displaystyle \lim _{T \rightarrow \infty } \frac{1}{\sqrt{T}} \int _{-T}^{T} d\tau \tilde w_l(\tau ) e^{-i\omega \tau },
\end{align}
and
\begin{align}
\label{Fourier2}
P_l ^* (\omega ) \equiv \displaystyle \lim _{T \rightarrow \infty } \frac{1}{\sqrt{T}} \int _{-T}^{T} d\tau P_l(\tau) e^{-i\omega \tau }.
\end{align}
The equation $\lim _{\omega \rightarrow 0}I_Z(\omega) =0$ is satisfied if
\begin{align}
\label{D=02}
\lim _{\omega \rightarrow 0} \tilde w_1 ^* (\omega ) = 0.
\end{align}
Equation $(\ref{D=02})$ holds if
\begin{align}
\label{D=0}
\nonumber E_l &\equiv \lim _{\omega \rightarrow 0} \tilde w_l ^* (\omega )
\\ &= 0,
\end{align}
which we will show in the rest of this section.

Let us consider the Fourier transform of both sides of Eq.~$(\ref{self-consistent1})$:
\begin{align}
\label{Fourier7}
\tilde w_l ^* (\omega ) \displaystyle = P_l^*(\omega ) + g^*(\omega ),
\end{align}
where $g^*(\omega )$
represents the Fourier transform of the last term of Eq.~$(\ref{self-consistent1})$.
Concerning the first term of the r.h.s. of Eq.~($\ref{Fourier7}$), condition $(\ref{criticaleq})$ yields
\begin{align}
\label{delta}
\lim _{\omega \rightarrow 0} P_l^*(\omega) = 0.
\end{align}
Namely, the first term in the r.h.s. of Eq.~$(\ref{self-consistent1})$ does not affect the value of $D$.

To proceed further,
we show that the kernel $A(t,t')$ is approximately represented in the form of $B( x t - yt')$ with certain constants $x$ and $y$ satisfying $xy>0$ if $K\approx K_c$ \cite{Daido}.
We divide $A$ as $A = A_{\mathrm{(e)}} + A_{\mathrm{(ne)}}$,
where $A_{\mathrm{(e)}}$ and $A_{\mathrm{(ne)}}$ represent the contributions from the entrained and nonentrained oscillators, respectively.
Because $\psi_j (t)$ of the entrained oscillators is constant,
$A_{\mathrm{(e)}}$ can be represented as follows:
\begin{align}
\label{entrained}
A_{\mathrm{(e)}} (t,t') = B_{\mathrm{(e)}}(t-t'),
\end{align}
where $B_{\mathrm{(e)}}$ is a certain function.
We can represent $A_{(\mathrm{ne})}$ as follows:
\begin{align}
\label{nonentrained1}
A_{\mathrm{(ne)}}(t,t') = \displaystyle \lim _{N\rightarrow \infty} N^{-1} \sum _j e^{2\pi i X_j(t,t')},
\end{align}
where
\begin{align}
\label{nonentrained2}
\nonumber \displaystyle X_k(t,t') \equiv & ( l \psi _k(t) - l' \psi _k (t') )
\\&- \frac{K}{2\pi i} \int _{t'} ^{t} d\tau \hat H'(\psi _k (\tau )).
\end{align}
In the case of a nonentrained oscillator, if we rewrite Eq.~$(\ref{nonentrained2})$ as
\begin{align}
\label{nonentrained3}
\nonumber X_k(t,t') = & ( l - l' ) \psi _k(0) + \Delta _{k}' ( lt - l't' )
\\ &+\hat X_k(t,t'),
\end{align}
where $\Delta _{k}'$ represents the mean frequency of $\psi _k$,
then the last term $\hat X_k(t,t')$ is bounded.
This is because the last term of Eq.~($\ref{nonentrained2}$) and $\psi_k (t)$ are periodic as shown in Appendix A.
Because the bounded variation of $\hat X_k(t,t')$ should be small compared to the other terms for $t\gg 1$,
we can neglect $\hat X_k(t,t')$ as follows \cite{Daido}:
\begin{align}
\label{nonentrained4}
X_k(t,t') = (l - l') \psi _k(0) + \Delta _{k}' (lt - l't').
\end{align}
This approximation is good enough near the critical point $K=K_c$ \cite{Daido} at which
the bounded function $\hat X_k(t,t')$ vanishes.
Therefore,
the kernel $A_{(\mathrm{ne})}$ is represented in the following form:
\begin{align}
\label{nonentrained5}
A_{(\mathrm{ne})} (t,t') = B_{(\mathrm{ne})}( l t - l' t'),
\end{align}
where
\begin{align}
\label{AAAA}
\nonumber \displaystyle &B_{(\mathrm{ne})}( l t - l' t')
\\&\equiv \lim _{N\rightarrow \infty } N^{-1} \sum _k \exp (2\pi i \{ (l - l') \psi _k(0) + \Delta _{k}' (lt - l't') \} ).
\end{align}
We can exclude the case of $ll'<0$ \cite{Daido}, in which, for large $t$ and $0\leq t^* \leq t$,
\begin{align}
\label{AAAAAAAA}
\nonumber \displaystyle &B_{(\mathrm{ne})}( l t - l' t^*)
\\\nonumber &= \lim _{N\rightarrow \infty } N^{-1} \sum _k \exp (2\pi i \{ (l - l') \psi _k(0) + \Delta _{k}' l(t + |(l'/l)| t^*) \} )
\\&= 0.
\end{align}
From Eqs.~($\ref{entrained}$) and ($\ref{nonentrained5}$), the kernel $A$ can be expressed by using a function $B$ as follows:
\begin{align}
\label{nonentrained6}
A(t,t') = B( x t - y t'),
\end{align}
where $x(l) y(l') > 0$.

In order to use the Fourier transform, we replace $\int _{0} ^t$ by $\int _{-\infty } ^{\infty}$ in the r.h.s. of Eq.~$(\ref{self-consistent1})$,
based on the discussion in Appendix B.
Then, from Eqs.~$(\ref{self-consistent1})$ and $(\ref{nonentrained6})$, we obtain
\begin{align}
\label{Fourier3}
\tilde w_l ^* (\omega ) \displaystyle = P_l^* (\omega )+ 2\pi i l K \sum _{l'} q_{l'}  \frac{1}{|x|}  B^* (\omega /x )  \tilde w_{l'} ^* (y \omega /x ),
\end{align}
where
\begin{align}
\label{Fourier5}
B^{*}(\omega ) \displaystyle \equiv \int _{-\infty }^{\infty } d\tau B(\tau ) e^{-i\omega \tau }.
\end{align}
Note that we have {\it not} divided the r.h.s. of this equation by $T$.
From Eqs.~$(\ref{delta})$-$(\ref{Fourier3})$, we can obtain
\begin{align}
\label{Fourier6}
\nonumber &\lim _{\omega \rightarrow 0} \tilde w_l ^* (\omega )
\\ &=\lim _{\omega \rightarrow 0} 2\pi i l K \sum _{l'} q_{l'}  \frac{1}{|x|}  B^{*} (\omega /x )  \tilde w_{l'} ^* ( y \omega /x).
\end{align}
Equation ($\ref{Fourier6}$) can be rewritten as follows:
\begin{align}
\label{Fourier9}
E _l = \sum _{l'} F_{l, l'} E_{l'},
\end{align}
where the coefficient $F_{l, l'}$ is a certain constant for $l= \pm 1,\pm 2,\cdots$ and $l'=\pm 1,\pm 2,\cdots$.
The trivial solution of Eq.~($\ref{Fourier9}$) is $E_l = 0$ for all $l$.
Let us assume that Eq.~($\ref{Fourier9}$) has another solution $E_l = \bar E_l$.
Then, we can easily show that $E_l = c\bar E_l$ with any constant $c$ is also a solution of Eq.~($\ref{Fourier9}$).
However, this statement contradicts the fact that $|E_l|^2$ (and $D$) is bounded for $K\not = K_c$.
Therefore, the only solution of Eq.~($\ref{Fourier6}$) must be $E_l = 0$ for all $l$.

In fact, in the limit $K \rightarrow K_c$, we can derive $E_l = 0$ for all $l$ as follows.
Let us divide $B^*(\omega)$ as $B^*(\omega) = B^{*}_{(\mathrm{e})}(\omega ) + B^{*}_{(\mathrm{ne})}(\omega )$,
where $B^{*}_{(\mathrm{e})}(\omega )$ and $B^{*}_{(\mathrm{ne})}(\omega )$ represent the contributions
from the entrained and nonentrained oscillators, respectively.
Condition $(\ref{criticaleq})$ yields $\lim _{\omega \rightarrow 0} B^*_{(\mathrm{ne})}(\omega ) = 0$.
In the limit $K \rightarrow K_c$, $B^*_{(\mathrm{e})}(\omega ) = 0$ because there are no entrained oscillators at $K=K_c$.
As a result, we obtain $E_l = 0$ for all $l$, because the right-hand sides of Eqs.~(\ref{Fourier6}) and (\ref{Fourier9}) vanish in the limit $K \rightarrow K_c$.
Consequently, Eq.~$(\ref{D=0})$ holds.

\section{Summary and Discussion}

We have investigated the statistical properties of long-term fluctuations in the system of globally coupled phase oscillators $(\ref{phasemodel})$ with general coupling,
by using the statistical quantity $D$,
which is the diffusion coefficient of the temporal integration of the order parameter.
To understand the finite size effects in the system behavior near the synchronization transition point, the scaling property of $D$ with system size $N$ has been examined.
We have demonstrated that $D\sim O(1/N^a)$ with a certain positive constant $a$ in the coherent regime, and $D\sim O(1)$ in the incoherent regime.
The difference in the scaling laws is caused by the difference in the correlations among the phases of the oscillators at different times;
these correlations remain after a long-term period in the coherent regime.
In other well-known systems such as the Ising model,
the correlation function of an order parameter decays exponentially with time \cite{Goldenfeldbook,Nishimori},
and thereby, $D$ follows $D\sim O(1)$ with respect to the system size $N$ both in the coherent and incoherent regimes except for the transition point.
For the phase oscillator model $(\ref{phasemodel})$,
such a difference in the scaling laws of $D$
has not been found for other statistical quantities such as the variance and the correlation time of the order parameter \cite{Daido}.
The scaling property of $D$ in the coherent regime has been further explored in the limit $N\rightarrow \infty$.
We have analytically demonstrated that $D=0$ in the limit $N\rightarrow\infty$ for the system with a wide range of general coupling functions. If the system exhibits periodic behavior,
this result would be trivial.
However, this is not the case because non-periodic (chaotic) behavior is present even in the coherent regime of the system,
as supported by a positive Lyapunov exponent \cite{Miritello}.
Although the finite size effects on the statistical properties in the phase oscillator model $(\ref{phasemodel})$
have been well studied for the sinusoidal coupling function \cite{Daido,Daido4,Pikovsky4,Hong,Hong3},
they have remained unclear for a general coupling function except for several properties \cite{Hildebrand,Buice}.
We have clarified one aspect of the finite size effects in the coherent state for coupling functions satisfying Eq.~$(\ref{criticaleq})$,
which holds for a large class of general coupling functions \cite{Daido3,Daido2}.

Our result is useful to derive the scaling property of $D$ for the coupling strength interval $|K-K_c|$.
From the scaling hypothesis \cite{Goldenfeldbook,Nishimori}, the variance and the correlation time of the order parameter are scaled as $V\sim |K-K_c|^{-\gamma}$ and $\tau\sim |K-K_c|^{-z}$, respectively, where $\gamma$ and $z$ are the critical exponents \cite{Goldenfeldbook,Nishimori}.
Combining these scaling laws and Eq.~$(\ref{DD})$,
we can derive the following scaling law:
\begin{eqnarray}
D &\sim & |K-K_c|^{-\gamma-z}  \int_{-\infty}^\infty f(s)ds.
\end{eqnarray}
From our numerical simulations, we found that, in the limit $N\rightarrow\infty$, $\int_{-\infty}^\infty f(s)ds$ goes to 0 in the coherent regime
whereas it is finite in the incoherent regime.
Therefore, if the bifurcation of the order parameter is supercritical, we obtain the following scaling law in the limit $N\rightarrow\infty$:
\begin{eqnarray}
D & \sim & \left\{ 
\begin{array}{ll}
(K_c-K)^{-\gamma-z} & {\rm for}\quad K < K_c, \\
0 & {\rm for}\quad K > K_c.
\end{array}
\right.
\end{eqnarray}
The critical exponent is dependent on $\gamma $ and $z$ in the incoherent regime,
whereas it is independent of them in the coherent regime.
It is known that $\gamma=z=1$ for the sinusoidal coupling function \cite{Daido,Hildebrand,Buice}.

\begin{figure}
\centering
\includegraphics[width=6cm,clip]{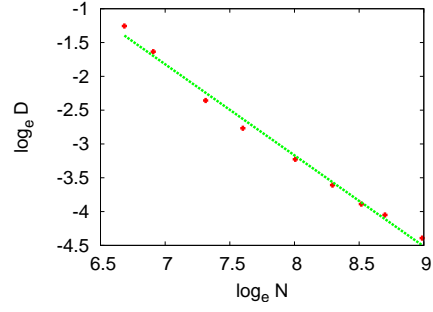}
\caption{The scaling property of the diffusion coefficient $D$ with system size $N$ in the coherent regime of the Kuramoto model,
where $K = 1.635 > K_c = 1.59\cdots$, $N= 800,\ldots, 8000$, and $\omega _j$ is deterministically generated.
The line fitting yields $D\sim N^{-1.347}$.
Each plot is an average over 10 different initial conditions.
}
\label{fig:DKdeter}
\end{figure}

There are two sources for the order parameter fluctuations.
The first is the oscillators that fail to synchronize with the order parameter motion.
The second is the randomness in the distribution of the natural frequencies.
In order to show that the main source of the fluctuations is the first one in the coherent regime,
we have performed numerical simulations by excluding the randomness of the natural frequencies.
Namely, the natural frequencies $\omega_j$ are not randomly but deterministically chosen from
the Gaussian distribution $G(\tilde{\omega})$ with mean zero and variance one,
i.e. $j/(N+1)=\int_{-\infty}^{\omega_j} G(\tilde{\omega})d\tilde{\omega}$.
Also in this case,
we have obtained the same scaling property of $D$ in the coherent regime, i.e. $D\sim O(1/N^a)$ with a positive constant $a$ for all the coupling schemes considered in this paper.
The result for the Kuramoto model is shown in Fig.~$\ref{fig:DKdeter}$.
Confirming the scaling law of $D$ in the incoherent regime should be our future work.

\section*{Acknowledgments}
We would like to thank K.~Ouchi,
Y.~Takahashi, and Y.~Sento for their fruitful discussions.
This research is supported by
Grant-in-Aid for Scientific Research (A) (20246026) from MEXT of Japan,
and by
the Aihara Innovative Mathematical
Modelling Project, the Japan Society for the Promotion of Science
(JSPS) through the ``Funding Program for World-Leading Innovative R\&D
on Science and Technology (FIRST Program)," initiated by the Council
for Science and Technology Policy (CSTP).

\appendix

\section{}

Because $\psi_k (t)$ is periodic, the last term of Eq.~$(\ref{nonentrained2})$ is also periodic as follows:
\begin{align}
\label{bounded}
\nonumber &\displaystyle \int d\tau \hat H'(\psi _k (\tau ))
\nonumber \\&=\int d\psi_k \frac{d\tau }{d\psi_k } \hat H'(\psi _k (\tau ))
\nonumber \\&= \int d\psi_k \frac{\hat H'(\psi_k )}{(\Delta _k - K \hat H (\psi_k ))} 
\nonumber \\&= - \frac{1}{K} \log |(\Delta _k - K \hat H (\psi_k ))|
\nonumber \\&= - \frac{1}{K} \log |d\psi_k (\tau ) /d\tau |.
\end{align}

\section{}

In order to use the Fourier transform,
we consider replacing $\int _0 ^t$ by $\int _{-\infty } ^{\infty}$ in the r.h.s. of Eq.~$(\ref{self-consistent1})$.
First, by defining $\tilde w_l(t) \equiv 0$ for $t<0$,
we can replace $\int _0 ^t$ by $\int _{-\infty } ^t$ in the r.h.s. of Eq.~$(\ref{self-consistent1})$.
For the group of entrained oscillators,
by defining $B_{\mathrm{(e)}}(t) \equiv 0$ for $t<0$,
we can further replace $\int _{-\infty } ^t$ by $\int _{-\infty } ^{\infty}$ in the r.h.s. of Eq.~$(\ref{self-consistent1})$.
For the group of nonentrained oscillators,
we separately treat the cases of $l=l'$ and $l \not= l'$.
In the case of $l = l'$,
we can replace $\int _{-\infty } ^t$ by $\int _{-\infty } ^{\infty}$ in the r.h.s. of Eq.~$(\ref{self-consistent1})$
by adequately defining $B_{\mathrm{(ne)}}(t) \equiv 0$ or $B_{\mathrm{(ne)}}(-t) \equiv 0$ for $t<0$ for each $l$.
In the case of $l \not = l'$, we replace $t'$ with a large $t^* (<t)$ in $B_{\mathrm{(ne)}}(t)$ of Eq.~($\ref{AAAA}$).
If $(l-l') \not = 0$ and both $t$ and $t^*$ are sufficiently large \cite{Daido}, we obtain
\begin{align}
\label{nonentrained7}
B_{\mathrm{(ne)}}(lt - l't^*) = \displaystyle &\lim _{N\rightarrow \infty} N^{-1} \sum _k \exp (2\pi i \{ (l - l') \psi _k(0) + \Delta _{k}' (lt - l't^*) \} )=0.
\end{align}
As a result, we can replace $\int _{-\infty } ^t$ by $\int _{-\infty } ^{\infty}$ in the r.h.s. of Eq.~$(\ref{self-consistent1})$.


\begin{thebibliography}{99}
\bibitem{assumption_pre}
If $Z(t)$ rotates around the origin of the complex plane for large $t$,
$w (t)$ may not be small.
However,
this situation is eliminated in our argument
because we take the limit $N \rightarrow  \infty $.
Note that
$\hat{Z} \equiv  \langle \lim _{N \rightarrow \infty} Z \rangle _t = \mathrm{const.} \not = 0$ in the coherent regime
(see p.29 in \cite{Crawford2}).
\bibitem{assumption}
If $Z_l(t)$ rotates around the origin of the complex plane for large $t$,
$w_l (t)$ may not be small.
However,
this situation is eliminated in our argument
because we take the limit $N \rightarrow  \infty $.
Note that
$\hat{Z_l} \equiv  \langle \lim _{N \rightarrow \infty} Z_l \rangle _t = \mathrm{const.} \not = 0$ in the coherent regime
(see p.29 in \cite{Crawford2}).
\bibitem{Fourier}
In the continuum limit of $N$,
both $\tilde w_l(t)$ and $P_l(t)$ have continuum spectra.
Therefore, Eqs.~($\ref{Fourier}$) and ($\ref{Fourier2}$) are divided by $\sqrt{T}$.
\end{thebibliography}
\end{document}